\documentclass{ar-1col-S2O}

\usepackage[comma]{natbib}
\usepackage{url}
\usepackage{lscape}

\begin{document}

\markboth{Ignas Snellen}{Exoplanet Atmospheres at High Spectral Resolution}

\title{Exoplanet Atmospheres at High Spectral Resolution}

\author{Ignas A. G. Snellen
\affil{Leiden Observatory, Leiden University, Postbus 9513, 2300 RA Leiden, The Netherlands}}

\begin{abstract}
High-resolution spectroscopy (HRS) has grown into one of the main techniques to characterise the atmospheres of extrasolar planets. High spectral resolving power allows for the efficient removal of telluric and host-star contamination. Combined with the large collecting area of ground-based telescopes it enables detailed studies of atmospheric species, temperature structure, atmospheric loss, and global winds and circulation patterns. In this review, the wide range of HRS observation and data-analysis techniques are described and literature results discussed. Key findings include:

\begin{itemize}

\item The highest irradiated planets show a rich spectrum of atomic 

and ionic species, just like stars. 

\item Retrieval analyses of Hot Jupiters and directly imaged Super-

Jupiters point to Solar metallicities and chemistry, but observed 

samples are still heterogeneous and incomplete. 

\item There appears to be a clear dichotomy between Hot Jupiters 

with and without atmospheric inversions, depending on their 

equilibrium temperature.

\item Some highly irradiated planets exhibit enormous leading and/or 

trailing tails of helium gas, providing unique insights into planet 

evolution and atmospheric escape processes.

\item Minor isotopes of carbon and oxygen are now being detected in 

gas giant planets and brown dwarfs with the interesting 

potential to shed light on formation pathways.

\end{itemize}

A list of potential pitfalls is provided for those new to the field, and synergies with JWST are discussed. HRS has a great future ahead with the advent of the extremely large telescopes, promising to bring temperate rocky exoplanets into view with their increase in HRS detection speed of up to three orders of magnitude.
\end{abstract}

\maketitle

\tableofcontents

\section{Introduction}
Since the discovery of the first extrasolar planets in the early 1990s, thousands of new worlds have been discovered. While the census of planetary systems is yet far from complete, it is clear that exoplanets are both common and diverse. They can be found in orbits of less than a day, but also at hundreds of astronomical units from their host star, and their masses and radii point to a broad continuum in compositions. Evidently, our Solar System is not a universal blueprint for other planetary systems. 

An important goal of exoplanet studies is to understand the formation and evolutionary processes driving this diversity. This will also enlighten us about the worlds in our own Solar System, e.g. explain why Earth and Venus are so different, or how planetary climates have evolved over time. Only by studying a wide variety of worlds, over a wide range of orbits, host-star properties, ages, and compositions, will we be able to identify families or classes of planets and understand their formation and evolutionary relationships.  This has also a strong philosophical component with a unique opportunity to learn about ourselves as highly-evolved organisms living on a temperate, water-rich world. How common are life-bearing planets? Are there any other worlds like Earth?

Some basic properties of exoplanets can be constrained using indirect detection methods, such as planet mass from radial velocity measurements, radius from transit observations, and mean density from the combination of the two. The latter can put constraints on a planet's bulk composition, but solutions are often highly degenerate and atmospheric measurements are required to gain more detailed knowledge on chemical make-up and local conditions. The first question that can be addressed is whether a planet has an atmosphere or not. Subsequently we want to know its chemical ingredients and temperature structure. Do we understand the physical and chemical processes in the context of its formation and evolution? In the search for habitable environments and extraterrestrial life we want to know whether there is evidence for water. Can we detect any potential biomarker gases, such as molecular oxygen, ozone, or methane? And ultimately, are these biomarkers indeed due to biological activity?

While space-based observations have kick-started the field of atmospheric characterisation more than two decades ago, ground-based high-resolution spectroscopy (HRS) has slowly but surely developed into a remarkable and unique way to probe exoplanets and forms the focus of this review. In the remainder of Section 1, a brief history of exoplanet atmospheric observations is presented, followed by an overview of the wide range of HRS observing methods. Section 2 provides technical details on HRS instrumentation, data acquisition, signal search-techniques, and retrieval analyses.  Section 3 discusses results for short-period planets that have been presented in the literature so far using HRS transmission and phase-curve measurements. Section 4 discusses these for directly-imaged planets for which HRS is combined with high-contrast imaging techniques.  Section 5 examines synergies with JWST observations, and Section 6 discusses the exciting future of HRS in the era of the Extremely Large Telescopes.

A graduate-level review on high-resolution spectroscopy for exoplanet atmospheres has been presented before by \cite{2018arXiv180604617B}. The general review on exoplanet atmospheres by \cite{2019ARA&A..57..617M} is also a great source of information. 

\subsection{A brief history of exoplanet atmospheric observations}

After the discovery of two planetary-mass objects orbiting a pulsar \citep{1992Natur.355..145W}, it was the detection of the first hot Jupiter by \cite{1995Natur.378..355M} that started a genuine scientific revolution. Significant skepticism existed in early days, since it was not generally accepted that the observed radial velocity variations were caused by orbiting planets \citep[e.g.][]{1997Natur.385..795G}. This resulted in a strong incentive to search for other evidence for exoplanets, in particular starlight scattered off their atmospheres \citep[][]{1999ApJ...522L.145C,1999Natur.402..751C}. Although unsuccessful, these attempts constitute the first time HRS was utilised to search for atmospheric signals. At high spectral resolution, reflected light can be disentangled from direct starlight due to its Doppler shift caused by the orbital motion of a planet. Unfortunately, hot Jupiters turn out to have very low albedos and no undisputed HRS scattered-light signals have been found so far. 

The groundbreaking discovery that the hot Jupiter HD 209458\,b transits its host star \citep[][]{2000ApJ...529L..45C} opened the way for a new technique: exoplanet transmission spectroscopy \citep[][]{2000ApJ...537..916S}. Subsequently, \cite{2002ApJ...568..377C} detected sodium in transmission using the HST STIS spectrograph, the start of a new subfield of exoplanet atmosphere measurements.  In the years that followed, space-observations with both HST and Spitzer Space Telescope were trail-blazing. \cite{2005ApJ...626..523C} and \cite{2005Natur.434..740D} presented the first secondary eclipse measurements of thermal emission using Spitzer, followed by the first infrared phase-curve and longitudinal heat-distribution \citep[][]{2007Natur.447..183K}. HST observations in ultra-violet light showed significant Ly$_\alpha$ and metallic transmission signals out to at least several planet radii, pointing to exospheric absorption and atmospheric loss processes \citep[][]{2003Natur.422..143V, 2004ApJ...604L..69V}. Absorption from water vapour turned out to be more elusive, and early claims \citep[e.g.][]{2007Natur.448..169T} were disputed, until water measurements became irrefutable with HST’s WFC3 instrument \citep[e.g.][]{2013ApJ...774...95D}. HST has been pushing the frontier of atmospheric characterisation up to the launch of JWST (see Section 5), by inferring hazes from hot-Jupiter spectra \citep{2016Natur.529...59S}, high-precision transmission spectroscopy of the mini-Neptune GJ 1214 \citep{2014Natur.505...69K} and cool planets such as K2-18b \citep{2019ApJ...887L..14B}, and many other results. 

In the meantime, theoretical studies showed the exciting possibilities of ground-based HRS transmission spectroscopy \citep[e.g.][]{2001ApJ...553.1006B}, but early attempts resulted in upper limits \citep[][]{2001ApJ...546.1068W,2001A&A...371..260M,2005ApJ...622.1149D}. The first successful ground-based observations targeted the sodium D lines in HD 189733\,b and HD 209458\,b using the Hobby-Eberly Telescope \citep[][]{2008ApJ...673L..87R} and Subaru Telescope \citep[][]{snellen_ground-based_2008} respectively, the latter confirming the earlier HST detection. While sodium has been detected in many Hot Jupiters since, recent observations of HD 209458 have cast some doubts on these early results \citep[][see Section 3.1]{casasayas-barris_atmosphere_2021}, mostly due to the treatment of stellar-related contaminants.   

Subsequently,  the CRIRES spectrograph on ESO’s Very Large Telescope was used to detect carbon monoxide in the transmission spectrum of HD 209458\,b \citep[][]{snellen_orbital_2010}, successfully utilising cross-correlation techniques for the first time. It demonstrated the unique capabilities of HRS compared to space-based low-resolution spectroscopy by the unambiguous detection of a molecular signal, and revealing Doppler effects of planet orbital motion and possible atmospheric winds. The change in the radial component of the orbital velocity of a planet was subsequently used to target thermal emission of the non-transiting planet $\tau$ Bootis\,b \citep[][]{brogi_signature_2012, 2012ApJ...753L..25R}, detecting carbon monoxide and revealing its orbital inclination, followed by the first ground-based detection of water in the thermal spectrum of HD 189733\,b \citep[][]{birkby_detection_2013}. 

An alternative route to characterise exoplanet atmospheres opened up through the first detections of substellar companions using high-contrast imaging \citep[HCI; e.g.][]{2004A&A...425L..29C,2008Sci...322.1348M,2010Sci...329...57L}. These Super-Jupiters are typically massive gas giants that straddle the planet/brown-dwarf divide residing on wide orbits in young ($<$100 Myr) star systems, and are almost solely in the observational realm of large ground-based telescopes with state-of-the-art adaptive optics systems.  Since every measurement directly probes such object,  atmospheric characterisation is immediate upon discovery and can be supplemented by observations at different wavelengths or high-contrast spectroscopy. In this way, the dominant spectroscopically-active species are readily observable in Super-Jupiters, such as water, carbon monoxide, and methane \citep[e.g.][]{2010ApJ...723..850B}, in addition to constraining the atmospheric temperature structure and possible presence of clouds. 
Since the spectroscopic signatures of stars and their planets are significantly different, many studies have recognised the power of combining high-contrast imaging with high-resolution spectroscopy \citep[e.g.][]{2002ApJ...578..543S,2007A&A...469..355R, 2014arXiv1409.5740K, snellen_combining_2015}. \cite{2013Sci...339.1398K} first demonstrated the strength of medium-resolution integral-field spectroscopy  on the HR8799 system with the 10m Keck OSIRIS spectrograph. Subsequently, CRIRES was used to target the massive gas-giant Beta Pictoris b, measuring the radial component of its orbital velocity and deriving its spin-rotation period via its $v\sin{i}$ \citep[][]{snellen_fast_2014}. 

Over the years, HRS has developed into a established observational technique for  exoplanet atmospheric characterisation.  In the current JWST era, which provides unprecedented and unrivalled capabilities in particular in the thermal infrared, ground-based HRS remains uniquely competitive and highly synergetic with space-based observations (see Section 5). E.g., HRS observations show that the optical spectra of the highest irradiated gas giants are dominated by neutral and ionised metals \citep[][]{hoeijmakers_spectral_2019}, with dayside spectra suggesting strong thermal inversions \citep{nugroho_detection_2020}. Atmospheric escape processes can now be probed very effectively at high spectral resolution using the helium 1083 nm line \citep[][]{2018Sci...362.1384A, nortmann_ground-based_2018}. A strong synergy is developing between spatially resolved HRS observations that reveal signatures of day-night winds,  zonal flows, and other atmospheric dynamics, and 3D atmospheric modelling \citep{louden_spatially_2015}. HRS spectroscopy of directly-imaged Super-Jupiters have allowed for the first isotope measurements \citep[][]{zhang_13co-rich_2021}. As discussed in Section 6, soon the Extremely Large Telescopes will further revolutionise the field of HRS atmospheric characterisation, ultimately opening up to temperate rocky planets.

\begin{figure}[t]
\vspace{0.5in}
\includegraphics[width=7.9in, angle=90]{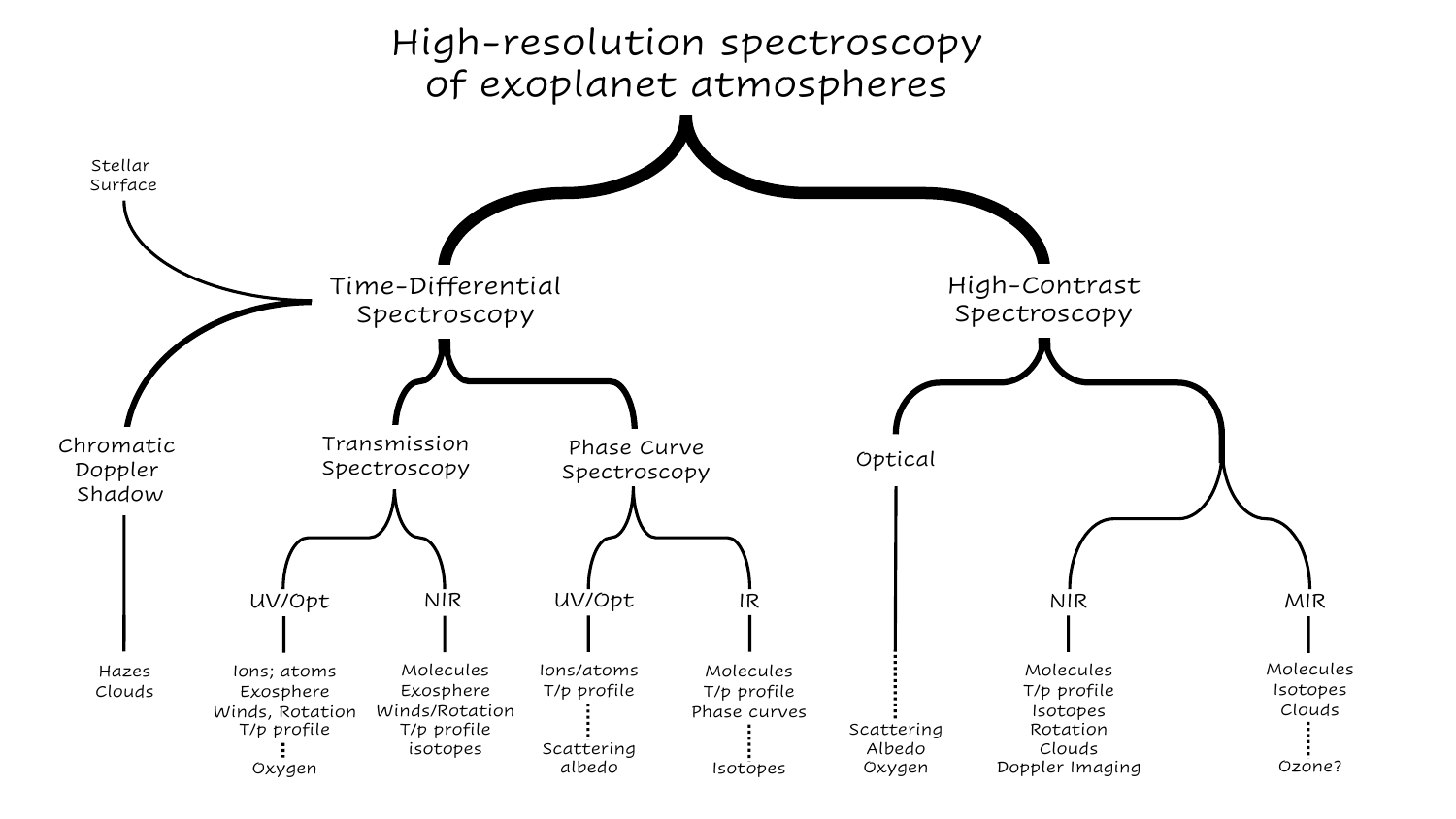}
\vspace{1.3in}
\caption{Diagram showing the range of methods employed to probe exoplanet atmosphere using high-resolution spectroscopy. The specific atmospheric properties that can be studied are indicated per wavelength regime. Dotted lines indicate future possibilities \citep[c.f.][]{2000RPPh...63.1209P}.}
\label{Fig1}
\end{figure}

\subsection{Overview of high-resolution spectroscopic methods}

Any atmospheric measurement requires radiation originating from the planet to be distinguished from that of the host star. Basically, two families of methods achieve this: either through time-differential observations where contributions from the planet change over time (e.g. during transit, eclipse or with phase-curve observations - technical details discussed in Section 2.2), or by angularly separating the planet from the star (e.g. direct, high-contrast imaging or interferometric observations). Figure \ref{Fig1} shows a diagram \citep[c.f.][]{2000RPPh...63.1209P} of the wide range of methods in which HRS is used to target exoplanet atmospheres, and which properties can be measured now or possibly in the future. 

Transmission spectroscopy measures the integrated light from the planet-star system during transit as function of wavelength \citep[e.g.][]{2019ARA&A..57..617M}. Absorption and scattering processes make the transparency of the atmosphere dependent on wavelength, and as such impact the effective planet size and transit depth. With HRS, the transmission spectrum is also sensitive to Doppler effects induced by the planet orbital motion and atmospheric dynamics such as day-to-nightside winds, equatorial jet-streams, or bulk-rotation. The transmission signal from a specific atmospheric species depends on its abundance as function of altitude in the atmosphere, the abundances of other species that may shield its signal, and the atmospheric temperature structure which is influenced by the surface gravity and mean molecular weight of the atmosphere. In the UV-optical, transmission signals generally come from atoms and ions originating from the hydrostatic part of the atmosphere or an extended exosphere that may indicate atmospheric escape. Transmission spectroscopy in the near-infrared is mainly sensitive to molecular absorption, but  exospheric helium can also be targeted. 

During transit, the Rossiter-McLaughlin effect (RM or Doppler shadow) induced by a planet, which is widely used to  measure spin-orbital alignments, can also be used to obtain a transmission spectrum \citep[][]{snellen_new_2004,dreizler_possibility_2009,2020A&A...644A..51S}. The amplitude of the RM effect (measured using stellar lines)  depends on the effective size of the planet, which is affected by atmospheric absorption and scattering as above.  By measuring chromatic variations in the RM-effect, broad-band transmission features can be determined, which are mostly inaccessible via classical HRS transmission spectroscopy.  During transit, information can also be obtained on the spectrum of the local stellar surface that is being transited, such as star spots and center-to-limb variations (CLV) in stellar line shapes and strengths \citep[][]{2016A&A...588A.127C}. The combined effects of the Doppler shadow, CLV, and planet atmospheric absorption can be very challenging to disentangle. 

Phase-curve observations are time-differential measurements away from the transit, during which the planet is seen from different viewing angles.  
In contrast to transmission spectroscopy, non-transiting planets can also be targeted. While in the latter case constraints on the planet radius are missing, the HRS phase-curve yields the orbital inclination of the planet and therefore the true mass. As for transmission spectroscopy, infrared phase-curve observations are dominated by molecular opacity sources.  
The temperature structure, i.e. temperature-pressure (T/p)  profile of the atmosphere, governs how molecular opacity is translated into the observed spectrum. An isothermal atmosphere produces (to first order) a black body spectrum without atomic or molecular lines. An atmosphere with an inverted or non-inverted T/p profile results in emission or absorption lines respectively. In the UV/optical the contrast between thermal emission from the planet and the star is generally very large. Only the hottest planets (i.e.  Ultra-Hot Jupiters; UHJ), can currently be studied in this way. They show mainly opacity from atoms and ions, but potentially also metal-oxides and metal-hydrides. For all but the hottest planets the UV/optical light will likely be dominated by star-light scattered off the planet, but no undisputed HRS detections have yet been presented in the literature. If the size of the planet is known (i.e. for transiting planets), scattered light will lead to the geometric albedo of the planet as function of wavelength, providing insights in scattering processes (Rayleigh, Mie) and possibly the scattering medium (clouds, hazes, molecules). Depending of the effective range of altitudes at which the starlight is being scattered, light travels back and forth through the planet atmosphere, but is also being absorbed, meaning that also the scattered light spectrum will contain information on atmospheric constituents. 

In the case of high-contrast spectroscopy (branch to the right of Fig. \ref{Fig1}), high-contrast imaging techniques are combined with HRS. Light from the star is removed as much as possible at the position of the planet to obtain its spectrum free from stellar contamination.  It requires an adaptive-optics-assisted HRS spectrograph, possibly in conjunction with coronagraphy. In principle, this technique can also be used in combination with interferometry, but no HRS modes yet exist for such instruments. So far, all directly-imaged planets amenable to HRS observations have more favourable star/planet contrasts because they are young and still warm from their formation. Since they are only detectable in relatively wide orbits, their spectra are completely dominated by thermal emission, and reflected starlight is negligible. As with phase-curve observations, their HRS spectra provide information on molecular species, atmospheric T/p profile, and the possible presence of clouds. Recently, the first isotopic measurements have also been obtained \citep[][]{zhang_13co-rich_2021}. Planet spin-rotation can be constrained using line broadening. Potentially, spatially resolved information of planets may be obtained, revealing possible dark or bright spots or bands, using Doppler imaging \citep[][]{crossfield_doppler_2014}. These will require highly accurate HRS spectral time series.

In the era of the Extremely Large Telescopes (ELTs), HRS instrumentation will improve sufficiently to target planets around our nearest neighbour stars in reflected light, providing constraints on albedo and atmospheric constituents at optical/NIR wavelengths. Arguably the holy grail in exoplanet science is the search for extraterrestrial life via biomarker gases. Molecular oxygen is potentially detectable in atmospheres of nearby rocky exoplanets either using HRS transmission spectroscopy or high-contrast spectroscopy \citep[e.g.][]{lovis_atmospheric_2017}. 

\begin{figure}[t]
\includegraphics[width=\textwidth]{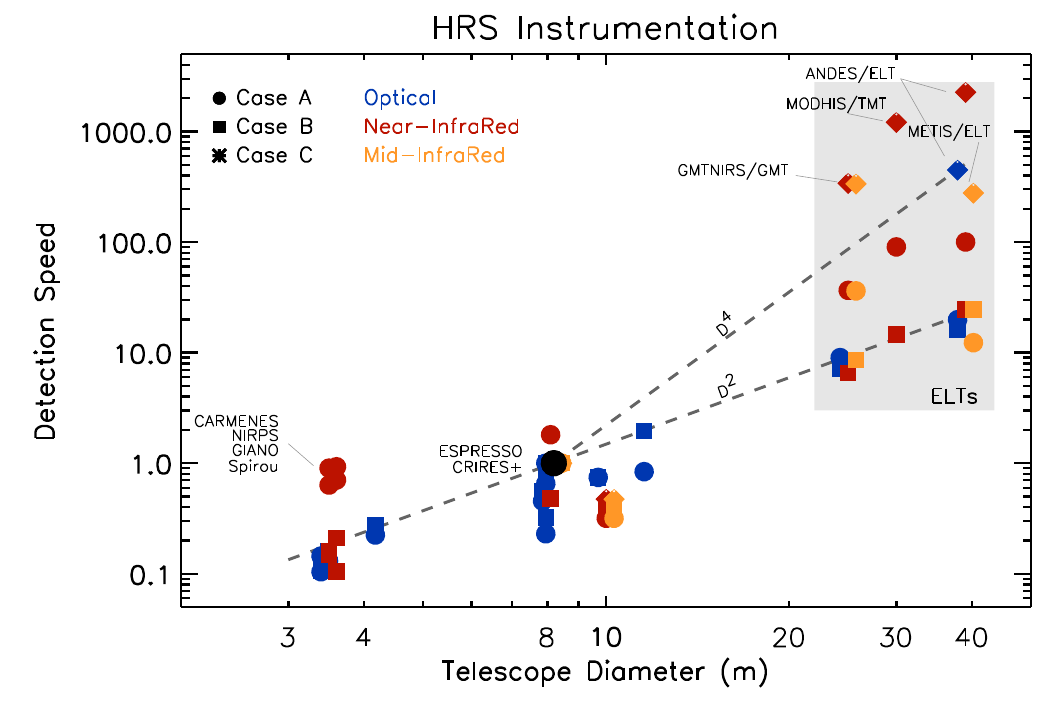}
\caption{Detection speed of HRS observations for current and planned HRS instrumentation, normalised to the ESPRESSO (optical) and CRIRES+ (IR) spectrographs on ESO's VLT. Three cases are specified (A, B, and C) as explained in the main text. ELT instrumentation will be up to three orders of magnitude faster than any instruments available now.}
\label{Fig2}
\end{figure}

\section{Detailed techniques}

In this section, HRS observing and data analysis techniques are described in detail. Firstly, the relevant properties of HRS instrumentation are discussed, how they influence detection speed, and how existing and planned HRS instruments compare to each other.  Section 2.2 reviews observation and data analysis strategies, Section 2.3 the many aspects of atmospheric modelling, and Section 2.4 provides a list of potential pitfalls. 

\subsection{HRS instrumentation}

Over the past decades, there has been an enormous incentive to develop stable, high-precision, high-resolution spectrographs for the search of extrasolar planets using the radial velocity technique. These same instruments are utilised for exoplanet atmospheric observations. Only in recent years have instruments been developed with atmospheric work as their main science driver \citep[e.g.][]{mawet_keck_2016,vigan_first_2024}. A key property is the detection speed, $\Gamma$, which governs the observing time needed to achieve a detection of a certain atmospheric species at a set signal-to-noise (S/n) ratio.  Here, three very simplified cases, A, B, and C, are discussed. 

In Case A, the target spectrum consists of a set of well-separated, infinitely narrow lines of equal strength. Assuming a telescope diameter $D$ and a telescope + instrument throughput of $f$, the detection speed is proportional to,

\begin{equation}
\Gamma_A \propto D^2 f  \lambda_{\rm{span}} R 
\end{equation} 

where $R$ is the resolving power and $\lambda_{\rm{span}}$ the instantaneous wavelength coverage of the spectrograph. Note that this is only a gross simplification of reality for many reasons, since this would imply that one should always push for the largest resolving power and wavelength coverage. First of all, $\lambda_{\rm{span}}$ and $R$ are generally not independent from each other due to the limited detector space. Also, the throughput $f$ of a spectrograph typically decreases with increasing $R$, which depends on its instrumental design, level of atmospheric turbulence, and the presence and performance of a possible adaptive-optics system. A higher $R$ generally makes telluric line removal (Section 2.6) more effective, and a minimum $R$ is needed to measure certain physical characteristics, such as thermal line broadening, spin rotation, or planet orbital velocity. In addition, a higher $R$ means that lines are less blended, increasing the overall HRS signal. A higher resolving power also means that possible stellar effects, such as center-to-limb variations and Doppler shadow can be disentangled from the planet signal more easily. On the other hand, for a certain $R$ the planet lines will start to be resolved, above which the detection speed will no longer increase. In reality, some wavelength regions will be richer in spectral features than others, depending on the type of planet. This will influence the effect of $\lambda_{\rm{span}}$ on the detection speed. 

In Case B, the target spectrum consists of effectively one single, infinitely narrow line, e.g. exospheric helium, H$\alpha$. This requires only a very limited instantaneous wavelength coverage (as will also be the case for molecular oxygen in future reflected light studies). In this case  the detection speed is proportional to

\begin{equation}
 \Gamma_B \propto D^2 f R
 \end{equation}
 
In Case C we consider high-contrast spectroscopy for which the residual seeing-halo of the host star is the dominant source of noise. Under such circumstances, the detection speed is proportional to,

\begin{equation}
\Gamma_C \propto D^4 f \lambda_{\rm{span}} R
 \end{equation}

This depends on the telescope diameter $D$ to the fourth power, because a change in $D$  influences the size of the seeing disk in terms angular resolution elements. In reality, however, this also critically depends on the performance of the adaptive optics (AO) system, the angular distance of the planet to the star in $\lambda/D$, and possible sky background and instrumental noise contributions. 

Optical high-resolution spectrographs have almost exclusively been designed for RV measurements that require extreme wavelength stability at sub-meter-per-second precision over multi-year timescales. This is often achieved with a combination of light-scrambling, a fibre feed, temperature/pressure stability, and simultaneous wavelength calibration. Since atmospheric signals are generally too weak to measure planet radial velocity at accuracies $<$100 m/s, such wavelength precision may be considered unnecessary. However,  a very stable spectrograph with minimal instrumental artefacts does significantly help in removing telluric and systematic effects, but may negatively affect its throughput. 

Several teams are developing instrumentation specifically aimed at combining high-resolution spectroscopy with high-contrast imaging, such as the Keck Planet Imager and Characteriser (KPIC) that combines starlight suppression using Keck's AO system and single-mode fibres transporting the light to the upgraded NIRSPEC with a spectral resolving power of R=35,000 \citep[][]{mawet_keck_2016,delorme_keck_2021}. Similar instrument designs are chosen for VLT/HiRISE \citep[][]{vigan_first_2024} and RISTRETTO \citep[][]{chazelas_ristretto_2020}, and Subaru/REACH \citep[][]{2020SPIE11448E..78K}. Interestingly, due to a reduced coupling in single-mode fibres, stellar flux is suppressed relative to that of the planet, a mechanism that is also utilised in the GRAVITY instrument at the VLT Interferometer \citep[][]{2020A&A...633A.110G}.

Table \ref{Table1} shows high-resolution spectrographs that are currently most frequently used for exoplanet atmospheric observations, and those that are in the planning for the upcoming ELTs. Fig. \ref{Fig2} shows their detection speed for the different cases. Since $\Gamma$ scales with $D^{2-4}$, the instruments on the largest telescopes are generally more powerful. Since optical high-resolution spectrographs have similar instantaneous wavelength coverages, spectrographs on 8-10m class telescopes are typically 4-6$\times$ faster than those on 4m-class telescopes (1 transit on a larger telescope equals 4 to 6 transits on a smaller one). In the near-infrared wavelength regime, this is true for some instruments on smaller telescopes (such as CARMENES, SPIRou, GIANO and NIRPS) partly compensated by their larger $\lambda_{\rm{span}}$ (Case A) and high throughput. 

The planned high-resolution spectrographs on future extremely large telescopes will be particularly powerful for {\sl Case C}, since the detection speed scales roughly with $D^4$, e.g. making METIS on the ELT almost 300$\times$ faster than CRIRES+, and ANDES at the ELT and MODHIS on the TMT more than a thousand times. Do note, however, that for the comparison of all these instruments it is assumed that throughput and adaptive optics performances are the same, which will be a challenge.

\begin{table}
\caption{High resolution spectrographs, currently most used or in planning. Columns from left to right show the telescope, its primary mirror diameter, name of the spectrograph, entrance type (slit or fibre), resolving power,  wavelength regime, its instantaneous wavelength coverage expressed as $\lambda_{\rm{span}}/\lambda$, and whether an adaptive optics system is present or not. The last three columns show the detection speed relative to ESPRESSO and CRIRES+ on ESO's VLT for the different cases, assuming identical throughput and AO performance, as explained in the main text and presented in Figure \ref{Fig2}.}
\label{Table1}
\begin{tabular}{llrcc|c|ccc}\\ \hline
Telescope (m) &Spectrograph & $R$ & $\lambda_{\rm{min}}-\lambda_{\rm{max}}$& $\frac{\lambda_{\rm{span}}}{\lambda}$& AO &\multicolumn{3}{c}{Detection Speed}  \\
                    &   (fibre or slit)                     & &   & && A & B & C \\ \hline
\multicolumn{3}{l}{{\sl Optical}}&&&&&&\\
VLT (8.2)   & ESPRESSO (f)  & 140k&0.38$-$0.79  & 0.70 &$-$ & \bf{1} & \bf{1} \\
                 &  UVES (s)           &  100k&0.30$-$1.10  & 0.64  & $-$&0.7&0.7\\
Keck (10)           & HIRES (s)      & 70k&0.30$-$1.00  & 0.70 & $-$ &0.8&0.8 \\
LBT (2x8.4) & PEPSI (f) & 130k&0.38$-$0.91  & 0.30&$-$&  0.8 &2.0 \\         
          Gemini (8.1)      & MAROON-X (f) & 80k&0.50$-$0.90  &0.57& $-$&0.5&0.6\\
Subaru (8.2) & HDS (s) & 45k&0.30$-$1.00  & 0.50 & $-$&0.2&0.3\\ 
DCT (4.3) & EXPRES (f) & 150k&0.38$-$0.68  & 0.57&$-$&0.2&0.3\\ 
ESO (3.6) & HARPS (f)& 120k&0.38$-$0.69  & 0.58&$-$&0.1 &0.2  \\
     CFHT (3.5)           & ESPaDOnS (f) & 81k&0.37$-$1.05  &0.96&$-$&0.1&0.1 \\
Calar A. (3.5) & CARMENES (f) &95k&0.52$-$0.96  &0.59&$-$& 0.1 & 0.1 \\
TNG (3.6) & HARPS-N (f) & 120k&0.38$-$0.68  & 0.58&$-$&0.1 &0.2 \\
ELT (39)  & ANDES (f) & 100k&0.40$-$1.00  & 0.86 &$+$ & 20 &16 &449 \\
GMT (25) & G-CLEF (f) & 100k&0.35$-$0.90  & 0.88&$-$& 9 & 7  \\   
\multicolumn{3}{l}{{\sl Near-Infrared}}&&&&&&\\
VLT (8.2)     & CRIRES+ (s)         &  100k&0.95$-$5.30  & 0.14 &  $+$& \bf{1} & \bf{1}& \bf{1}\\
Keck (10)& NIRSPEC (s) & 35k&0.95$-$5.50 & 0.11 & + & 0.3 & 0.4 & 0.5\\                  
Gemini (8.1)  & IGRINS  (s)          &   45k&1.45$-$2.50   & 0.53 &$-$ &1.8 & 0.5 \\
Calar A. (3.5)& CARMENES (f) &80k&0.96$-$1.71  &0.56&$-$& 0.6 & 0.2 \\
               ESO(3.6)           & NIRPS (f)&  82k&0.95$-$1.80  & 0.62&$-$ &0.7 & 0.1 \\
                CFHT(3.5) & Spirou (f) & 75k&0.95$-$2.35 & 0.85&$-$& 0.9&0.1\\
TNG (3.6)  & GIANO (f) & 50k&0.90$-$2.50  & 0.94&$-$&0.7&0.1\\ 
ELT (39)     & ANDES (f) & 100k&1.00$-$1.80  & 0.57 &$+$ & 100 &25 &2264 \\
TMT (30) & MODHIS (f) & 100k&0.95$-$2.40  & 0.87 &$+$ &90&15&1210\\
GMT (25)           & GMTNIRS (f)  & 50k &1.07$-$2.45   & 0.78  &$+$ &37 & 6 & 340                     \\     
\multicolumn{3}{l}{{\sl Mid-Infrared}}&&&&&&\\
VLT (8.2)     & CRIRES+  (s)       &  100k&0.95$-$5.30  & 0.14 &  $+$& \bf{1} & \bf{1}& \bf{1}\\
Keck (10)& NIRSPEC (s)  & 25k&0.95$-$5.50 & 0.11 & + & 0.3 & 0.4 & 0.5  \\                  
ELT (39) & METIS (s) & 100k&3.00$-$5.00  & 0.07 &$+$ & 12 &25 &278\\
GMT (25)      & GMTNIRS (f) & 100k &2.90$-$5.30   & 0.59 &$+$ &   36& 9 &   336              \\       \hline                          
\end{tabular}
\end{table}

\subsection{Observation and data analysis strategies}

Here we explain how HRS observations are typically performed, and discuss procedures for data analyses. For transmission and phase curve observations we review signal-search techniques, and for the former how stellar surface effects like the Doppler shadow and center-to-limb variations can influence observations and need to be accounted for.  

\subsubsection{Transmission and phase curve observations}

\begin{figure}[t]
\includegraphics[width=\textwidth]{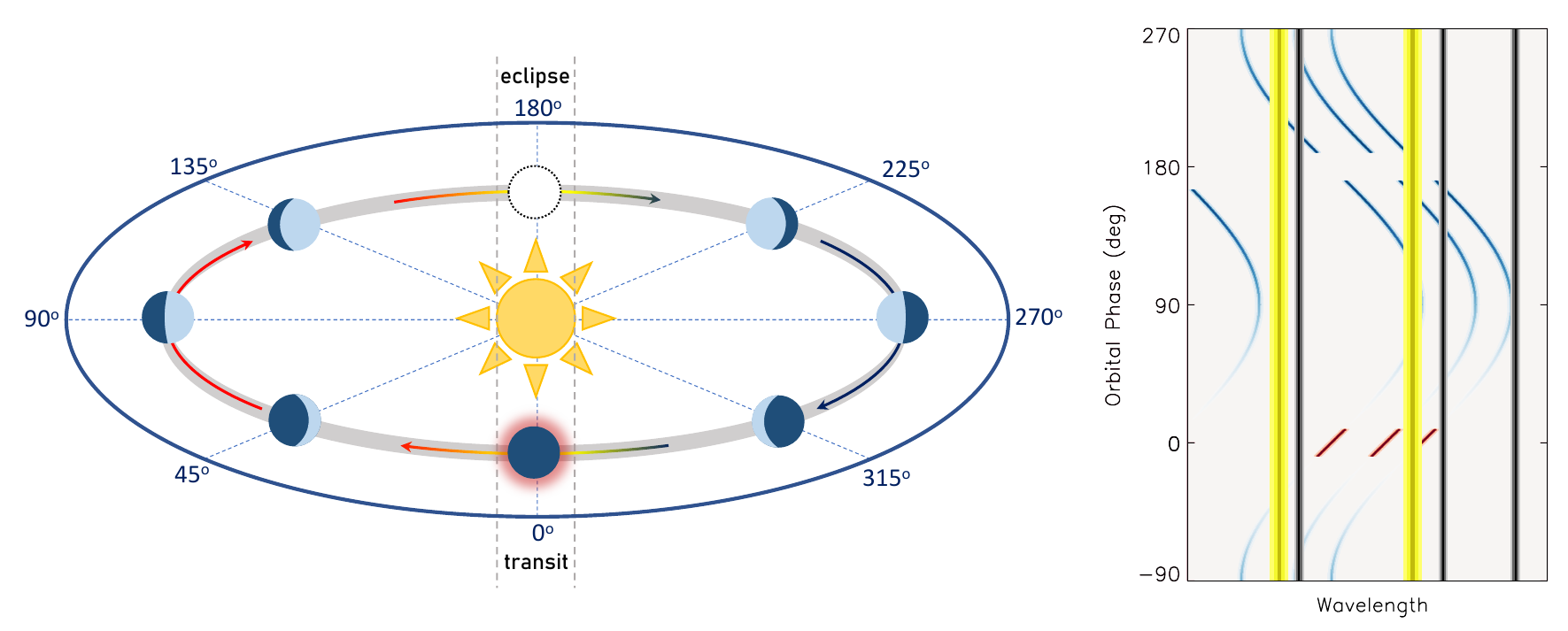}
\caption{Appearance of a planet as function of orbital phase (left). If in a near edge-on orbit the system will exhibit transits, allowing transmission spectroscopy, and eclipses - when the planet is blocked from view.  The diagram on the right shows how in an observing sequence telluric (black) and stellar (yellow) lines are quasi static in wavelength, while those features originating from the planet (dayside emission in blue, transmission in red) exhibit strong Doppler effects. }
\label{Fig3}
\end{figure}

A vital feature of time-differential HRS observations is their sensitivity to changes in Doppler-shift of the planet spectrum as function of orbital phase, as shown in Figure \ref{Fig3}. The radial component of the planet orbital velocity can vary by tens of km sec$^{-1}$ over a night's observations. Planet lines therefore rapidly change position relative to the much stronger stellar and telluric features. This means that the latter can be removed efficiently with a minimal impact on the planet signal. For this to work best, instrumental stability is key. Therefore, most observations consist of a continuous series of exposures where possible calibrators are only invoked before and/or after the main observing sequence. Secondary eclipses are generally avoided since no exoplanet signal is present during these times and planet-free spectra are often not needed for calibration purposes. HRS phase curve observations are therefore also well feasible for non-transiting planets. 

When planning observations, one should avoid individual exposures that are too long such that planet signals get spectrally blurred. E.g., a planet in a circular 1-day orbit around a solar-type star has an orbital velocity of $\sim$200 km sec$^{-1}$, with a change in the observed radial component of up to 1 km sec$^{-1}$ per minute. While up to now most HRS transit and phase curve observations are limited to hot Jupiters with fast-changing Doppler shifts, planets in wider orbits and/or around lower-mass stars have significantly lower orbital velocities, causing the change in Doppler shift to be insufficient to be utilised for telluric and spectral line removal. However, successful HRS phase-curve observations near quadrature, and transit measurements on relatively broad spectral features, indicate that  HRS spectroscopy is also possible under these circumstances using other calibration methods, e.g. by comparing in and out of transit data \citep[][]{cheverall_feasibility_2024}, or spectra taken at very different phases. 
 
\paragraph{Analysis of transmission and phase-curve data} 

The first important element of the data analysis of transmission and phase-curve data is the removal of the stellar and telluric contaminations. The telescope and instrument throughput generally varies with time and wavelength, e.g. due to seeing and airmass effects, which are impossible to calibrate accurately. To mitigate this, spectra are normalised and often high-pass filtered such that any broadband information is lost. To first order, the stellar lines do not vary in strength and position with time, meaning that removal of the mean spectrum will take care of the stellar contribution. While the telluric line positions are quasi-stable with time, the line strengths vary due to changes in airmass and possible changes in the observatory's atmospheric circumstances. To first order, these variations are similar for all tellurics and therefore can be measured in some strong lines and applied to all others. If needed, such correction can be performed for different telluric species, e.g. water and methane, separately. 

For high signal-to-noise data, however, there are more subtle issues. The wavelength solution of the instrument may vary over time, or Earth-atmospheric winds may change the exact wavelength position of telluric lines. Seeing effects may change the spectral resolution for slit spectrographs. Stellar lines move due to the Earth's rotation and gravitational pull of the planet overnight, and may change in shape and strength due to stellar activity. Instead of aiming to model all these different effects, a blind search for systematic effects in the data can be achieved using principle component analysis. Currently, one of the most popular algorithm is SysRem \citep[][]{2005MNRAS.356.1466T}, originally developed to detrend lightcurves of transit surveys. SysRem and other algorithms that are being used to remove telluric and stellar contamination, all will affect and remove part of the target planet spectrum, an effect that ideally needs to be taken into account e.g. by analysing artificial injection signals (see below). In future analyses, it may be possible to simultaneously forward model and retrieve the stellar spectrum, telluric contamination, instrumental effects and the planet spectrum without losses. Such approach is already being implemented in high-contrast spectroscopy \citep[][]{2019AJ....158..200R} for which the ratio of planet signal to the various noise contributions is significantly more favourable. 

\paragraph{Cross-correlation, signal-search techniques, and detection significance}

The procedure to remove telluric and stellar features as described above will largely preserve the line-features in planetary spectra because of the change in Doppler shift of the planet over the observing sequence. Except for single-line analyses, the next step is to optimally combine the signals from different lines in the spectra, with each individual line still buried in the noise. This is often done by cross-correlating the residual spectra with a model template \citep[see for a detailed description the review by][]{2018arXiv180604617B}. The more similar the model template is to the planet  spectra, the higher the cross-correlation signals will be. In the ideal case, the cross-correlation signal of a spectrum with $N$ well-separated, equally strong lines will boost the signal-to-noise by a factor of $\sqrt{N}$ compared to that of one individual line. Alternative methods exist to extract the total planet signal, such as techniques using Doppler tomography \citep[][]{2019MNRAS.490.1991W,2024MNRAS.531.3800M}, and log-likelihood methods that directly compare model templates with the data \citep[][]{gibson_detection_2020}. 

In the case of direct fitting of template models to spectra, the change in planet Doppler shift as a function of time can be included as extra free parameters. Assuming circular orbits, and assuming that the orbital period and time of inferior conjunction are known, the radial velocity of the planet depends on the RV amplitude, $K_p$, and the system velocity, $V_{sys}$, where the former is governed by the stellar mass and orbital inclination. In the case of combining cross-correlation signals, these are often calculated, Doppler shifted, and added, for a wide range of ($K_p$,$V_{sys}$) combinations to produce a so called $K_p$-$V_{sys}$ diagram. 
This has the added benefit that such a diagram provides an independent estimate of the significance and reliability of any detected signal by showing the cross-correlation noise properties in regions of the $K_{\rm{p}}$-$V_{\rm{sys}}$ diagram where no signal is expected (see below). Note that atmospheric circulation patterns such as day-to-nightside winds may alter the exact $K_{\rm{p}}$ and $V_{\rm{sys}}$ of the planet signal. Non-circular orbits make the $K_{\rm{p}}$,V$_{\rm{sys}}$ diagram less intuitive \citep[e.g.][]{2024A&A...688A.191G}.

\begin{figure}[t]
\includegraphics[width=\textwidth]{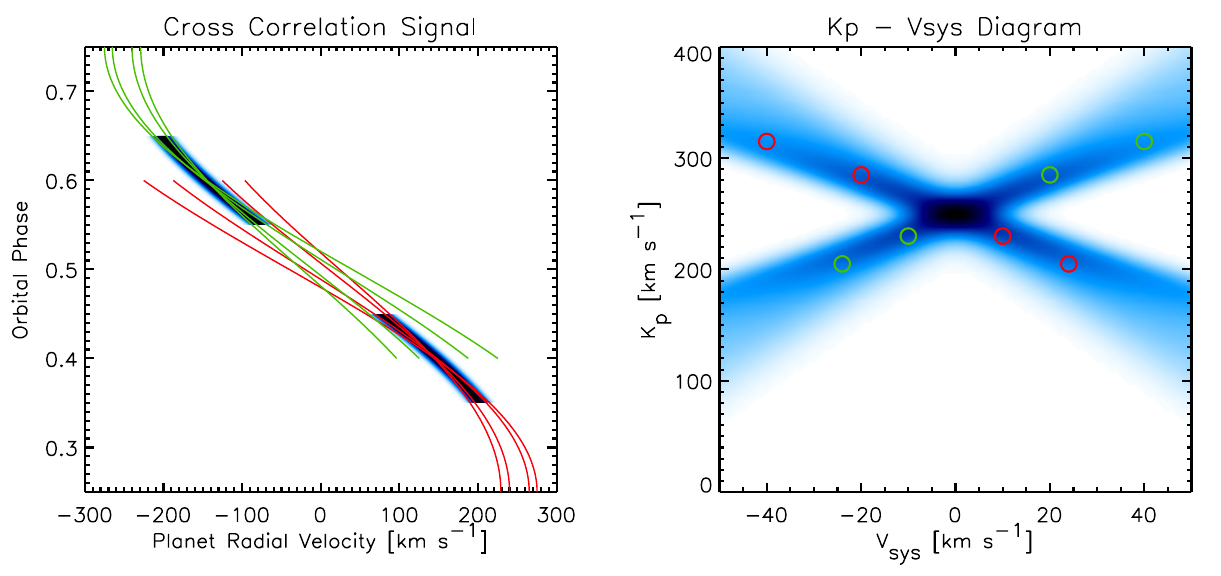}
\caption{The left panel shows a simulation of a cross-correlation signal (in blue-black) of a hot Jupiter observed over two orbital phase ranges, before ($\phi \sim$0.4) and after ($\phi \sim$0.6) secondary eclipse. The right panel shows the resulting $K_{\rm{p}}-V_{\rm{sys}}$ diagram with its characteristic X-shaped appearance. The origin of this X-shape comes from the range of ($K_{\rm{p}}$,$V_{\rm{sys}}$) combinations that fit either the pre-eclipse or post-eclipse part. This is shown by the red and green orbital solutions in the left panel, which correspond to the ($K_{\rm{p}}$,$V_{\rm{sys}}$) combinations indicated by the red and green circles in the diagram on the right.}
\label{Fig4}
\end{figure}

The data points in a $K_{\rm{p}}-V_{\rm{sys}}$ diagram are not independent from each other. Figure \ref{Fig4} shows such diagram for a noiseless time-series of cross-correlation signals (left panel) of a hypothetical hot Jupiter and the corresponding $K_{\rm{p}}-V_{\rm{sys}}$ diagram (right panel). A signal from any single observation can in principle be fitted with a linear combination $K_{\rm{p}}$ and $V_{\rm{sys}}$. For this reason, observations taken over a relatively short orbital phase range result in characteristic patterns as seen in the right panel of Fig. \ref{Fig4}, with a typical X-shape if observations are combined from both before and after the secondary eclipse. 

The $K_{\rm{p}}-V_{\rm{sys}}$ diagram is often used to estimate the statistical significance and reliability of an observed signal by comparing it to the noise properties in this diagram away from the expected $K_{\rm{p}}$ and $V_{\rm{sys}}$. One should be careful that the noise properties are a function of $K_p$, since the lower the $K_p$, the smaller the change in radial velocity during the observing sequence and the more PCA algorithms will both remove the noise and any planet signal. For strong signals, aliasing may affect noise estimates. Alternatively, the Welch T-test can be used on the cross-correlation data by comparing the distribution of cross-correlation (CC) values inside and outside the expected radial-velocity trail (e.g. as shown in the left panel of Fig. \ref{Fig4}). One should take into account that neighbouring CC values may not be independent of each other and that the properties of the CC-noise may not be Gaussian, two basic assumptions of the Welch T-test. 

HRS observers often make use of artificially injected signals. They can make sure that the different elements of the analysis pipeline work properly, they can help to assess the quality of the observed data throughout the night and/or over multiple nights and identify in which Echelle spectral orders specific molecules exhibit the strongest signals or where (residual) tellurics increase the local noise.

\paragraph{RM Effect, Doppler Shadow, and CLV}

In the case that HRS observers pursue a transmission spectrum with exoplanet atmospheric lines that are also present in the photosphere of its host star, the subsequent data analysis is prone to become significantly more complicated. Examples are alkali metals such as sodium in Hot Jupiters, and metal lines in Ultra-Hot Jupiters (UHJ, T$_{\rm{eff}} > 2000$ K). As a consequence, two effects may interfere with the planet transmission spectrum: the planet Doppler shadow and stellar center-to-limb variations (CLV). 

The Doppler shadow is inherently the same as the Rossiter-McLaughlin (RM) effect and constitutes a temporal deformation of rotationally-broadened stellar line profiles during transit. The occulted part of the stellar surface  momentarily does not contribute to the overall stellar spectrum, leaving a 'shadow' in the rotationally-broadened lines. In the course of a transit, this shadow moves from the blue to the red wing of a stellar line, depending on the spin-orbit angle and the transit impact parameter. The effective strength of the shadow depends on the opacity of the stellar line and the size of the planet. The Doppler shadow, which has a sign that is opposite to any exoplanet transmission signature, has to be modelled and removed from the data so not to bias the results. In most systems the transmission signal and Doppler shadow cross-over in velocity at some point during transit. At such a moment in time, the effective size of the planet is larger due to the additional absorption in its atmosphere, increasing the strength of the shadow making it very hard to disentangle the two effects. Data from such an epoch are therefore often masked. 
 
Center-to-limb variations (CLV) are also an important effect \citep[][]{czesla_center--limb_2015,yan_effect_2017}. The local spectrum near the stellar limb is significantly different from that near the centre of the stellar disk due to the different angle of incidence. It means that near the beginning or end of a planet transit, the spectrum of the occulted stellar surface can be significantly different from the average spectrum and could be falsely attributed to a planet transmission signal. 

In contrast to CLV, whose effects can be modelled and removed, other stellar surface inhomogeneities are more challenging. A planet's path over the stellar surface may cross more or fewer star spots, i.e. the occulted stellar surface may have on average a different effective temperature than the unocculted stellar surface. Such effects currently plague JWST transmission spectroscopy of planets around cool dwarf stars, unable to differentiate between atmospheric transmission features and stellar surface effects \citep[e.g. for water;][]{2023ApJ...948L..11M}. HRS observations could potentially help to distinguish between the two, because the radial velocity of the occulted stellar surface is mostly different from that of the planet - although effects are small.
 
\subsubsection{High-contrast spectroscopy}

High-contrast HRS relies on a telescope with an adaptive optics (AO) system and optionally a coronagraph, from which light can be fed to a high-resolution spectrograph. Only in the special cases for which the planet is sufficiently far away from its host star, is no AO-system required. In the case of planet thermal emission, the planet spectrum is in most cases significantly different from that of the star, because the effective temperature of the planet is much lower. In the case of scattered light, the planet and star spectrum are quasi-identical, but the reflected light spectrum for the planet will be Doppler shifted for most orbital phases. Therefore, in both cases, planet light can be spatially as well as spectroscopically separated from that of the star, and many studies have identified high-resolution spectroscopy in combination with high-contrast imaging as a potentially powerful method to find new planets and characterise them \citep[e.g.][]{2002ApJ...578..543S,2007A&A...469..355R, 2014arXiv1409.5740K, snellen_combining_2015}.

Blind searches for extrasolar planets with unknown positions require a form of integral field spectroscopy, which is currently only available at moderate spectral resolution (such as OSIRIS on the Keck Telescope, and ERIS and MUSE on ESO's VLT). In the case of a slit or fibre spectrograph, the planet position relative to that of the star must be known at the time of observation. If the host star is positioned on the detector and no coronagraph is present, exposure times are generally short to avoid saturation, meaning that the fainter planet spectrum could be degraded by instrumental read-out noise. On the other hand, the inclusion of a bright star spectrum can be beneficial for the data analysis.

The observed spectrum at the planet position generally contains flux from the star and planet, both modified by telluric absorption, and possibly a sky background contribution. The latter can be effectively removed by moving the target along the slit or IFU in between exposures. To first order, the stellar contribution (including tellurics) at the planet position will only slowly vary with wavelength, allowing it to be scaled and removed or included in the forward modelling, with the observed star spectrum as a concurrent reference. The residual spectrum can subsequently be searched for planet molecular signatures using cross correlation techniques as for transit and phase-curve observations. However, a particularly high data-quality requires more complex data analyses.  E.g. in the case of slit-spectroscopy a combination of good seeing and AO performance can cause the effective size of the star to be smaller than the slit-width. This can cause the planet-spectrum and associated telluric absorption to have a higher spectral resolution than the star-spectrum with associated tellurics. Also, the wavelength solution may vary along the slit. Effective strategies for the removal of the stellar and telluric contributions need to take this into account and will depend on the particular atmospheric circumstances and planetary system. E.g., these can include radiative transfer modelling of telluric absorption and a set of principal components of the residuals to include systematic effects \citep[][]{2023AJ....165..113R}. Ideally, the stellar, planet, telluric contributions and resolution effects should be simultaneously modelled to retrieve reliable and unbiased atmospheric parameters and their uncertainties, but this is beyond current methods and computer capabilities. 

\subsection{Atmospheric modelling}

Atmospheric modelling is an integral part of HRS observations. Model templates are used to tease out planetary signals, but also for atmospheric retrieval studies. Any HRS observation can only be as good as the models used. Equally important: any model is only as good as its underlying opacity data. Small errors in line positions can already lead to significant losses in signal. Therefore line positions need to be accurate to a fraction of the spectral resolving power. Another challenge is the line strengths, which become increasingly difficult to calculate and measure in laboratory settings at higher temperatures. Pressure broadening calculations may also be relevant, in particular for thermal emission spectra. Here we discuss the origin of opacity data, common software packages for radiative transfer calculations, and review the art of atmospheric retrievals. 

\subsubsection{Atomic and molecular opacity data}

Accurate line data for atomic and ionic species are mostly readily available from databases such as from the National Institute of Standards and Technology (NIST\footnote{https://www.nist.gov/pml/atomic-spectra-database}) and the Vienna Atomic Line Database (VALT\footnote{https://vald.astro.uu.se/}). However, the spectra of polyatomic molecules are extremely complex with potentially up to billions of transitions playing a role. While a proper discussion of the challenging practicalities of generating molecular line lists is beyond the scope of this review, this is a complex industry that deserves more attention. E.g. it took a significant part of a quantum-chemistry PhD project to calculate the line data for phosphine up to T=1500 K \citep[][]{2015MNRAS.446.2337S}. In addition to accurate line position, line broadening calculations can also be important for retrieval analyses.

A leading supplier of molecular line data at high temperatures is the ExoMol Group, working with the concept of {\sl first principles theory informed by experiment} \citep[][]{2021A&G....62.6.16T}. ExoMol now validates and adjusts line energies depending on laboratory measurements, leading to very high accuracy line lists. Its database contains line lists for more than 80 molecules, often including isotopic variants. High temperature line lists have also been produced by the TheoRets project \citep[][]{2016JMoSp.327..138R}, MoLLIST \citep[][]{2020JQSRT.24006687B}, and by NASA Ames. The HITRAN database \citep[][]{2017JQSRT.203....3G} provides line data applicable at low temperatures for several dozens of small molecules present in the Earth atmosphere, and expanded to higher temperatures in HITEMP \citep{2020ApJS..247...55H}. If the line data are insufficiently accurate or incomplete it can significantly hamper the detection of molecules, such as shown by \cite{hoeijmakers_search_2015} for the case of TiO and 
\cite{de_regt_quantitative_2022} for VO. The optical spectra of both these molecules are particularly challenging as they are dominated by electronic transitions, but significantly improved line data have become available since \citep[][]{2019MNRAS.488.2836M,2024MolPh.12255299B}.

\subsubsection{Radiative transfer calculations}

Radiation transfer calculations are needed to produce model exoplanet emission and transmission spectra. An example of an easy-to-use publicly available python package for this purpose is petitRADTRANS \citep[][]{2019A&A...627A..67M}, but there are many alternatives such as HELIOS \citep{2017AJ....153...56M}, EXo-REM \citep{2015A&A...582A..83B}, NEMESIS \citep{2008JQSRT.109.1136I}, HyDRA \citep{2018MNRAS.474..271G}, CHIMERA \citep{2014ApJ...793...33L}, PLATON \citep{2019PASP..131c4501Z}, and others. These codes typically require pre-computed opacity grids of relevant atoms, ions and/or molecules as function of pressure and temperature, in addition to continuum opacity sources such as $\rm{H}_2-\rm{H}_2$, $\rm{H}_2-\rm{He}$ collision-induced absorption and H$^-$. 
Most radiative transfer codes have also a way to include the effects of clouds, either physically motivated (e.g. MgSiO$_3$, Fe, etc.) or simple grey cloud models, controlled by parameters such as the atmospheric pressure and opacity at the cloud base, and a power law index that sets the decay towards lower pressures.

For the calculation of transmission spectra, in its basic form, the planet atmospheric structure is assumed to be independent of longitude or latitude, and the assumed pressure-temperature profile and surface gravity (e.g. at 1 bar) is converted to a radius-temperature relation and parameterised as a set of layers, each with a specific pressure and temperature. The optical depth for a ray of light grazing the atmosphere is determined as a function of impact parameter, and subsequently converted into a transmission spectrum.  For the calculation of planet emission spectra, similar set-ups are used for calculating the transmission of each layer, using a range of angles between the direction of the light rays and the atmospheric normal to determine the average emission from the planet disk. Since the full radiative transfer solution is a scattering atmosphere is computationally slow, it is often assumed that the scattering is weak compared to the absorption.

\subsubsection{Retrievals of atmospheric parameters}

The aim of atmospheric retrievals is to determine atmospheric properties and their uncertainties from an observed spectrum. In the context of exoplanets, this was first pioneered by \cite{2009ApJ...707...24M}, forward modelling millions of (low resolution) spectra for a ten-dimensional grid of input parameters. Subsequently, for all these model spectra it was determined whether they, within uncertainties, fit the observed data to derive the atmospheric properties.

Atmospheric retrievals for HRS are challenging for several reasons. In the case of HRS transit and phase-curve spectroscopy the observed spectrum is modified by the data reduction processes, such as PCA and filtering techniques. The forward models need to go through similar data processing to allow for a reliable comparison with the data. Furthermore, HRS data volumes can be enormous with up to tens of thousands of pixels per individual spectrum. 
\cite{brogi_retrieving_2019} presented a Bayesian atmospheric retrieval framework applicable to HRS observations, for which they developed a mapping from cross-correlation to log-likelihood. 
\cite{2020MNRAS.493.2215G} subsequently introduced a framework that directly computes the likelihood of a model fit to observations. Monte Carlo Markov Chains (MCMC) and Nested Sampling techniques can then be employed to explore parameter space. \cite{fisher_interpreting_2020} have presented a random forest supervised machine learning technique to perform the retrievals. \cite{2019AJ....158..200R} pioneered an approach in which the signal is forward modelled in such a way that  pre-processing of the data (such as cross-correlation and data interpolation) is minimised, preserving the statistical properties of the data as much as possible and therefore the likelihood accuracy. 

A key aspect of atmospheric retrievals is the choice of assumptions for chemical modelling and parameterisation of the atmospheric temperature structure.  When chemical equilibrium is assumed,  chemical abundances vary as function of altitude and strictly depend on temperature, pressure, metallicity and atomic ratios such as C/O, while disequilibrium processes such as photochemistry or rain-out condensation are ignored. Since vertical mixing time scales can be significantly shorter than those to reach chemical equilibrium, in particular at low pressures, a quench pressure can be introduced \citep{2014ApJ...797...41Z}. 
 Below this pressure, higher up in the atmosphere, the chemical abundances are assumed to be governed by chemical equilibrium at the quench pressure because vertical mixing dominates. E.g. in some hot Jupiters the temperature at mbar pressures is low enough that significant methane would form if the atmosphere would be in chemical equilibrium. However, the gas largely originates from deeper atmospheric layers with higher temperatures and does not have time to adjust to the local circumstances \citep[][]{2006ApJ...649.1048C}.
An alternative assumption is to adopt {\sl free chemistry} without any restrictions on volume mixing ratios except that they are constant as function of altitude. The abundance of every absorber is then a free, independent parameter.
If the quality of the data allows, it can be very informative to compare outcomes of different retrieval set-ups to address the validity of chosen assumptions. 
 
 Also the atmospheric temperature-pressure profile can be parameterised in many ways, from a complete free format, to largely physics-informed T/p profiles \citep[e.g.][]{2010A&A...520A..27G}. A choice will also have to be made whether to include clouds (physically motivated or grey) in the retrievals or not.  If an opaque cloud layer is present, but clouds are not included in the retrieval set-up, the retrieved T/p profile may get deformed. E.g. since in such case no flux is coming from below the cloud layer, this can be mimicked by an isothermal profile \citep{2017MNRAS.470.1177B}.     

\subsection{Potential pitfalls}

HRS can be a very complex technique, both in terms of observations and data analysis. Here a non-exhaustive list of potential pitfalls is discussed for those HRS users new in the field. 

\paragraph*{1. Exposure time blending} When planning transit or phase curve spectroscopic observations, individual exposure times should not be set too long to avoid spectral blending. The change in the radial velocity of the planet may be so large during the exposure that its affects the effective resolving power of the observations.

\paragraph*{2. Overlap of transmission signal with Doppler shadow} For some transiting systems, the trail of the Doppler shadow largely overlaps with that of the planet transmission signal. This means that at any moment during transit, the radial component of the orbital velocity of the planet matches the radial velocity of the occulted stellar surface, making analyses of planet transmission features that are also present in the stellar photosphere very challenging. Notable examples are HD 209458\,b and MASCARA 1\,b. 

\paragraph*{3. Velocity broadening of scattered light signals} When probing starlight reflected off atmospheres of short-period planets, one has to take into consideration that reflected signals can be significantly velocity broadened \citep[e.g.][]{spring_black_2022}. While the spin-velocity of the host star may be very slow, in the rest frame of the planet it will make about one revolution per orbital period, making detection of scattered light more challenging. Since in the tau Bootis system both star and planet are tidally locked, the scattered stellar lines are 'unbroadened' and expected to be significantly more narrow than those in the direct stellar spectrum, making it a popular target for reflected light studies \citep[][]{1999ApJ...522L.145C,1999Natur.402..751C,hoeijmakers_searching_2018}, but the planet turned out to be very dark.

\paragraph*{4. Conversion of signal-to-noise to statistical significance} The signal-to-noise ratio of a measurement can be converted into a statistical significance by assuming a Gaussian noise distribution. Since in this case a signal at S/n=3 corresponds to a 0.3$\%$ probability that it is caused by a random noise fluctuation, this could in principle be considered as a firm detection.  However, the fact that to arrive at such detection many post processing choices have been made should not be overlooked; how to extract the spectra and remove the telluric and stellar absorption lines, which spectral model to choose to perform the cross-correlation, and freedom in the exact location of the signal in the (K$_{\rm{p}}$, V$_{\rm{sys}}$) diagram due to a priori unknown atmospheric winds effects. Also, overall it can be difficult to assess to what extent the noise distribution is indeed Gaussian. These effects make the probability that a spurious signal is seen at S/n=3 significantly higher than 0.3\%. Therefore, many studies use a threshold of S/n=4 or 5 to claim a detection, which still may not be sufficient. Even more conservative is to require a detection in multiple transits or phase curve observations.

\paragraph*{5. Accidental fabrication of fake signals} For many HRS measurements, data over multiple nights and/or many spectral orders have to be combined. This is often done by deriving the CC functions per order and night which are subsequently combined using optimal weights. 
 These weights can be important because the data quality may be different from one night over the other, and a specific molecule may only have significant opacity in certain spectral orders of the spectrograph. It is of  upmost importance that the calculation of these weights is purely based on injected signals and not on the observed data. Otherwise the analysis may result in a fake, fabricated signal \citep[see for a discussion;][]{cabot_robustness_2019}. A modern high-dispersion spectrograph, for example, can have 80 spectral orders. If there is no planetary signal  at all, and the resulting cross-correlations from the individual orders contain all pure uncorrelated Gaussian noise, still $\sim16\%$ and $\sim$2.5\% of the orders will exhibit a 1$\sigma$ and 2$\sigma$ CC-peak respectively, which with optimal weights will combine into a completely spurious signal of $\sim$6$\sigma$. Note that the contribution from the signal injection to the cross-correlation function needs to be separated from that of the observed spectrum, otherwise the latter is still prone to bias the determination of the optimal weights \citep{cheverall_robustness_2023}. 

Also, the number of PCA or SysRem components, required to remove the telluric and stellar features in the spectra, needs to be chosen, which may also depend on the observing night and spectral order (e.g. depending on the strength of the telluric features). These need to be treated in the same way. 
 
\paragraph*{6. Unrealistic and biased uncertainties in retrieved parameters} Results from atmospheric retrieval analyses are mostly not free from biases and can provide unrealistically small uncertainties. This can have several reasons which are important to investigate. Different data sets, i.e. with different wavelength regimes, spectroscopic resolving power, or S/n levels, can result in retrieval results that are significantly different from each other. 
First of all, it is generally assumed that spectral data points and their uncertainties are independent of each other, which is often not the case, e.g. due to model inaccuracies or errors in telluric removal. The atmospheric temperature structure (and abundance profiles), i.e, can be parameterised in many ways, possibly none providing the perfect solution and therefore all resulting in their own biases. One way to characterise the strength and scale-length of correlated noise is using Gaussian Processes and include them in the retrieval set-up \citep{kawahara_autodifferentiable_2022}. In addition, uncertainties from pre-processing are mostly not included, such as small errors in the flux calibration from a spectroscopic standard star that  result in inaccuracies in the spectral slope. These can be included by adding extra order-normalisation parameters, spectral slope parameters, or by removing any low-frequency spectral information (by high-pass filtering) altogether. 

Another important source of unrealistically small uncertainties is the inclusion of too few free parameters in the retrievals. The Bayesian Information Criterion (BIC) is widely used to assess whether extra parameters provide a significantly better fit. If not, such extra parameters are not included in the retrieval. However, even if extra parameters do not result in a better fit, they may still influence the uncertainty (and best-fit value) of other parameters. For example, for an observed spectrum with relatively low S/n, 
retrievals may suggest that there is no statistical evidence for any deviation from chemical equilibrium, or for the presence of clouds. However, performing a retrieval with free-chemistry and with including the possibility of clouds, or 3D effects, can significantly change the outcome and enlarge the uncertainties on retrieved abundances. Sometimes strict assumptions are chosen, because the data would otherwise not provide meaningful constraints. 

\section{Results on close-in exoplanets} 

In this section, literature results on HRS atmospheric characterisation of close-in exoplanets, mostly hot Jupiters, are summarised and discussed. In such a vibrant field of research as exoplanet characterisation it is impossible to provide a complete overview of all the work presented in the literature, in particular because during writing new studies appeared on an almost daily basis. We review observations of specific atmospheric species, constraints on temperature structure, atmospheric dynamics and outflows, reflected light searches, and chromatic Doppler-shadow measurements.

\subsection{Measurements of atoms, ions, and molecules}

\paragraph*{Sodium H \& K lines}

Sodium was predicted to produce the strongest features in optical transmission spectra of Hot Jupiters \citep{2000ApJ...537..916S} and was the first chemical element detected in an exoplanet atmosphere, in HD 209458\,b using the HST \citep{2002ApJ...568..377C}. Studying sodium in transmission is challenging because stellar sodium lines are very strong, in particular for G and K-type stars. This means that the signal-to-noise of a planetary sodium signal at mid-transit is generally low and Doppler-shadow and CLV effects are particularly strong. While \citet{snellen_ground-based_2008} confirmed the original HST detection of sodium with Subaru, \citet{casasayas-barris_atmosphere_2021} find no evidence for sodium at this level in superior ESPRESSO observations taken more than a decade later. For this system, the radial velocity trail of the planet almost overlaps with the Doppler shadow, making the analysis of this target particularly difficult. In the earlier studies, CLV effects were not taken into account, so at least part of the original sodium signal could be attributed to this. On the other hand, due to the overlap of the RV-trail and Doppler shadow, the RM effect at sodium wavelengths is stronger in the case of a planetary atmospheric contribution, counteracting a transmission signal \citep[see][for a recent discussion]{2024A&A...683A..63C}.  Alternatively, long-term variability (i.e. due to changes in cloud-coverage in the terminator region) could play a role, although temporal variability can too easily be invoked to explain differences between data sets. 

\begin{figure}[t]
\includegraphics[width=\textwidth]{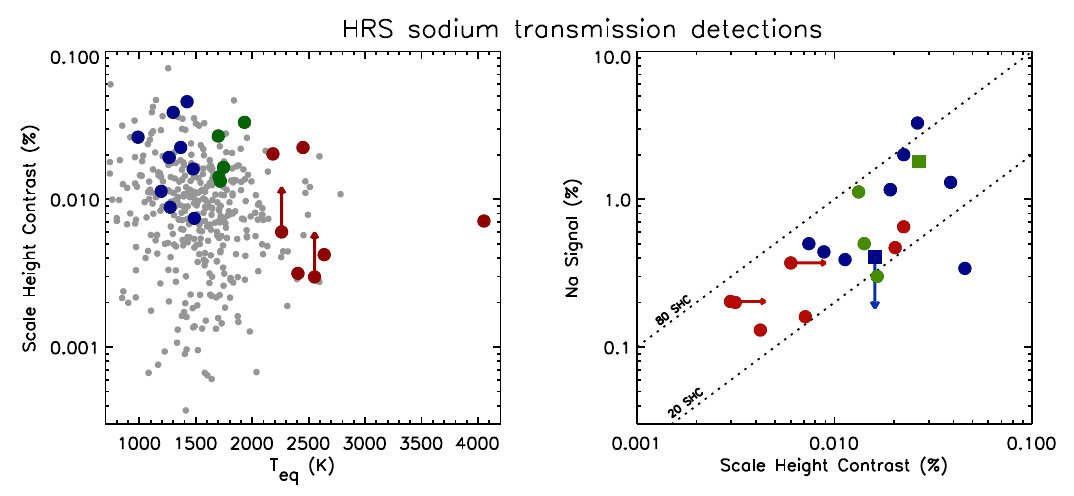}
\caption{Left panel: Currently known population of transiting Hot Jupiters around bright stars$^2$ with orbital periods $<$10 days, masses $>$0.1 M$_{\rm{Jup}}$, and radii $> 0.5 R_{\rm{Jup}}$. Their equilibrium temperatures (assuming a zero Bond albedo) are plotted versus Scale Height Contrast (see main text).  The coloured symbols (red: $T_{\rm{eq}} >2000$ K; green:  $1500\ \rm{K}\ <T_{\rm{eq}} <2000$ K; blue: $T_{\rm{eq}} <1500$ K) are  planets with HRS sodium detections presented in the literature. Those with lower limits in their scale height contrast have yet an unknown mass. The signals of the two planets (HD209458\,b and WASP-17\,b) marked with squared symbols are derived from their integrated light curve, assuming the sodium signal is 0.5\AA\ wide. Right panel: Strength of the sodium signal versus Scale Height Contrast. The data, as well as literature references, are provided in Supplementary Table 1.}
\label{sodium_fig}
\end{figure}

The left panel of Figure \ref{sodium_fig} provides an overview of the currently known population of transiting Hot Jupiters around bright stars ($V<13.5$) available in the NASA Exoplanet Archive\footnote{retrieved on 4/1/2024 from https://exoplanetarchive.ipac.caltech.edu}. Plotted are their equilibrium temperature, T$_{\rm{eq}}$, versus what we define as the Scale Height Contrast (SHC). The latter is the expected transmission signal from an opaque atmospheric ring with a thickness of one scale height, $H$, corresponding to 
\[ SHC = (2 \pi  R_p H)/ (\pi R_s^2)\] 
where $R_p$ and $R_s$ are the planet and host star radius respectively. It is a first-order measure of how strong a transmission signal is expected to be. 
The thick circles indicate those Hot Jupiters that have an HRS sodium detection presented in the literature, where the colour is an indication of T$_{\rm{eq}}$. HRS detections exist for hot Jupiters with a wide range of equilibrium temperatures, up to the hottest planets known.
This literature sample of planets with sodium detections is likely marred with biases. Also non-detections and their associated upper limits are poorly reported and are not included here. The SHC of those planets with detections is larger than average, probably because mostly promising targets are being observed, and those with a smaller SHC fall below typical detection limits. 
 
 The right panel of Figure \ref{sodium_fig} shows the observed strength of the sodium signals versus SHC. The data, and their origin from the literature, are presented in Supplementary Table 1. This strength is simply taken as the relative depth of the transmission signal, ignoring spectral resolution and intrinsic width of the sodium features. Despite the many caveats regarding sample selection, completeness and other biases, it is interesting to see that the strengths of the sodium signals correlate with SHC. A similar analysis was performed by \cite{Langeveld_survey_2022} showing a comparable relation on a smaller sample. The sodium signals in Hot Jupiters extend typically between 20 and 80 scale heights above their continuum absorption. A notable exception is WASP-127\,b, which has the largest SHC in the sample but a relatively weak sodium feature \citep{allart_wasp-127b_2020}. Assuming that these sodium features dominantly originate from the hydrostatic part of Hot Jupiter atmospheres, the observed planet-to-planet variations could be driven by variations in high-altitude cloud decks, sodium abundance, velocity broadening, and photo-ionisation for planets at the highest temperatures. Alternatively, one could argue that many of the signals are so strong, associated with micro-bar pressure levels, that they likely originate from exospheric outflows \citep[][]{seidel_wind_2020}.

\begin{figure}[h]
\includegraphics[width=5in]{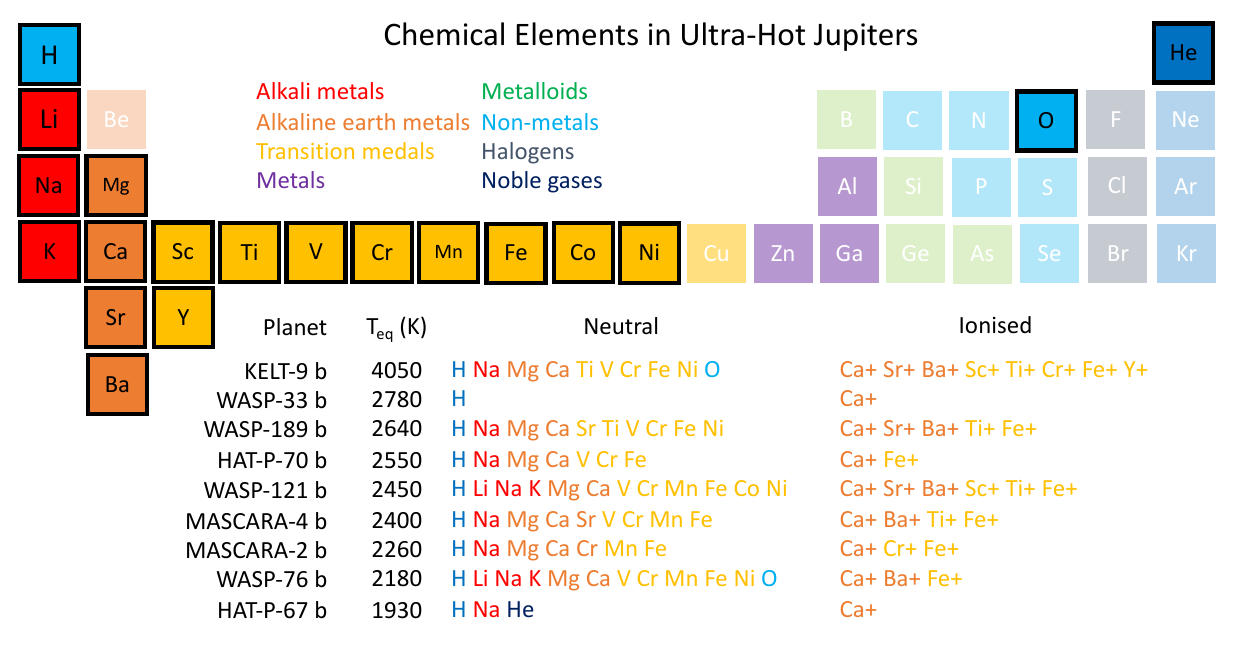}
\caption{Overview of the chemical elements detected in the transmission spectra of Ultra-Hot Jupiters (UHJ). Planets are ordered by decreasing equilibrium temperature. WASP-33\,b is a more challenging target due to the presence of significant stellar pulsations. References: \citet{borsa_high-resolution_2021}; \citet{borsato_mantis_2023}; \citet{hoeijmakers_atomic_2018, hoeijmakers_spectral_2019, hoeijmakers_high-resolution_2020, hoeijmakers_mantis_2022}; \citet{yan_extended_2018}; \citet{cauley_time-resolved_2021}; \citet{yan_ionized_2019};  \citet{prinoth_atlas_2024}; \citet{prinoth_time-resolved_2023}; \citet{stangret_detection_2020,stangret_high-resolution_2022};\citet{bello-arufe_mining_2022, bello-arufe_transmission_2023}; \citet{azevedo_silva_detection_2022}; \citet{ben-yami_neutral_2020};  \citet{merritt_inventory_2021}; \citet{maguire_high-resolution_2023}; \citet{gandhi_retrieval_2023}; \citet{jiang_detection_2023}; \citet{zhang_transmission_2022}; \citet{casasayas-barris_atmospheric_2019}; \citet{fossati_gaps_2023}; \citet{nugroho_searching_2020}; \citet{ehrenreich_nightside_2020}; \citet{tabernero_espresso_2021}; \citet{deibert_detection_2021}; \citet{pelletier_vanadium_2023}.}
\label{chemicals}
\end{figure}

\paragraph*{Other metals and ions}

Since sodium is so prominent in the optical transmission spectrum of Hot Jupiters, it is the subject of many studies. In particular in Ultra-Hot Jupiters (UHJ), numerous other metals and ions have been detected. \cite{2000ApJ...537..916S} already showed that potassium should also provide observable transmission signals for hot Jupiters. However, observational successes have been limited. While some popular  high-resolution spectrographs do not cover the  
potassium lines, they are at wavelengths hampered by telluric absorption from molecular oxygen. Initial reports of potassium from HST observations \citep{2015MNRAS.446.2428S} of WASP-31\,b have not been confirmed with HRS \citep{2019MNRAS.482..606G}. \citet{chen_detection_2020-1} seems the only reported HRS potassium detection, in WASP-52\,b. 

At temperatures T$_{eq} > 2000$ K, optical transmission spectra of gas giant planets are expected be cloud-free, close to chemical equilibrium, and dominated by features from atomic and ionised metals \citep{2018ApJ...863..183K}. Indeed, HRS has been prolific in detecting such signals in UHJs \citep{hoeijmakers_atomic_2018,hoeijmakers_spectral_2019}. Figure \ref{chemicals} shows a schematic overview of detections presented in the literature so far in the context of the periodic table, either in neutral or ionised state, including alkali metals (lithium, sodium, potassium), alkaline earth metals (magnesium, calcium, strontium, and barium), and transition metals (scandium, titanium, vanadium, chromium, manganese, iron, cobalt, nickel, yttrium). 

It is interesting to discuss those elements that are {\sl not} seen. Most of these have either  too low cosmological abundances, are insufficiently spectroscopically active, or have no significant lines in the optical wavelength regime  \citep[e.g.][]{kesseli_atomic_2022}. Also, condensation on the nightside can play a role, where those elements with a high condensation temperature (refractory elements such as Ti, Al, and Sc) could rain-out on the night side and be "cold-trapped" deep in the atmosphere, preventing them from being mixed back into the upper layers, resulting in low abundances in both dayside and transmission spectra \citep{hoeijmakers_mantis_2022,pelletier_vanadium_2023}. Titanium is indeed not detected in the three studied UHJs with the lowest equilibrium temperature (Fig. \ref{chemicals}). At lower temperatures it is expected that iron will rain out \citep{2018ApJ...863..183K, ehrenreich_nightside_2020}. This is in agreement with non-detections of iron and nickel in HAT-P-67\,b in contrast to sodium and ionised calcium \citep{bello-arufe_transmission_2023}.

Ionisation may also play a role in the apparent under-abundance or non-detections of neutral species \citep{kesseli_atomic_2022, gandhi_retrieval_2023}. Some metalloids and non-metals have been detected in dayside spectra \citep{cont_silicon_2022} or locked up in molecules \citep[such as CO;][]{yan_crires_2023}.  Detections of chemicals with very high atomic mass have been reported, such as rubidium and samarium \citep{jiang_detection_2023} in MASCARA-4\,b and terbium in KELT-9\,b \citep{borsato_mantis_2023}.  These species have very low cosmic abundances and their presence in UHJ atmospheres needs confirmation.

In recent years, great advances have been made in retrieval studies of exoplanet atmospheres such that abundances can be quantified. UHJs are ideal targets for this purpose due to the expected chemical equilibrium and absence of clouds.  Relative chemical abundances are generally much better constrained by transmission spectra than absolute abundances because they are less sensitive to the atmospheric temperature structure and absolute pressures probed. \citet{gibson_relative_2022}  retrieves relative abundances for WASP-121\,b consistent with solar values, except magnesium. For WASP-76\,b, \citet{pelletier_vanadium_2023} show seven species to have abundances relative to iron similar to proto-solar (except for those elements that are cold-trapped), while its overall metallicity appears only possibly slightly enriched (at $\sim1\sigma$) compared to its host star. \citet{gandhi_retrieval_2023}, in a homogeneous study,  retrieves the absolute and relative abundances from the optical transmission spectra of six UHJs. They show that iron abundances agree well with solar values, but that other species exhibit more variations for reasons discussed above. Note that these retrieval studies assume the atmospheres to be in hydrostatic equilibrium, while  actually some species may be associated with exospheric outflows \citep[][]{zhang_transmission_2022}.

\paragraph*{Carbon monoxide, water, and other molecules}

The first overtone absorption band of CO at 2.3 $\mu$m is an ideal target for ground-based HRS, due to its strong, well-separated set of lines in a relatively telluric-free part of the spectrum. Detections of CO and H$_2$O have historically dominated the HRS literature, shown to be present in both the transmission and dayside spectra of the two archetype hot Jupiters HD 209458\,b \citep{snellen_orbital_2010, brogi_framework_2017,hawker_evidence_2018} and HD 189733\,b \citep{de_kok_detection_2013, rodler_detection_2013, birkby_detection_2013, brogi_rotation_2016}. These molecular gases have so far not been widely studied in transmission spectra
 \citep{basilicata_gaps_2024,boucher_co_2023, maimone_detecting_2022, hood_atmospherix_2024}, in contrast to thermal dayside spectra, where they are found in relatively cool planets such as HD 189733\,b, 51 Pegasi\,b, to Ultra-Hot Jupiters \citep[e.g.][]{birkby_discovery_2017,brogi_roasting_2023,holmberg_first_2022}. In the case of non-transiting planets, detection of CO and/or H$_2$O have lead to the determination of their orbital inclinations, breaking the $\sin{i}$ degeneracy and solving for their planetary masses \citep{brogi_signature_2012,brogi_detection_2013,brogi_carbon_2014,rodler_weighing_2012,guilluy_exoplanet_2019}; a unique quality of high-resolution spectroscopy. 

Assuming that in atmospheres at T$>$1000 K, carbon and oxygen atoms are primarily locked up in CO and H$_2$O (ignoring CH$_4$ that can become abundant in carbon-rich atmospheres), atmospheric retrieval analyses can constrain the C/O ratios, which proposedly can shed light on planetary formation processes \citep{2011ApJ...743L..16O,2023ARA&A..61..287O}. However, care has to be taken in comparing results between different planets, since different assumptions in the retrieval set-ups (such as free chemistry, chemical equilibrium, or the presence of clouds) may bias the results. One could argue that the fact that most results are close to the solar C/O value (0.55) indicates that HRS atmospheric retrieval analyses are rather reliable. Interestingly, some C/O ratios derived from JWST observations deviate significantly from solar \citep[i.e. C/O$\sim$0.1 for HD 209458\,b;][]{xue_jwst_2024}. 
\citet{ramkumar_high-resolution_2023} find a C/O ratio consistent with solar for the UHJ MASCARA-1\,b, as do \cite{line_solar_2021} for WASP-77A\,b (C/O=0.59$\pm$0.08), while \citet{bazinet_sub-solar_2024} find a slightly elevated value for HIP 65A\,b. 
\citet{brogi_roasting_2023} and \citet{2023A&A...678A..23L} derive C/O values consistent with 0.75 for  WASP-18\,b and WASP 43\,b respectively, while the non-detection of H$_2$O in tau\,Bootis\,b points to a C/O$>$0.6 \citep{pelletier_where_2021}. However, by accounting for undetectable oxygen due to thermal dissociation \citet{brogi_roasting_2023}  derive C/O$<$0.34.

Other molecules than CO and H$_2$O are more rarely detected with HRS. The hydroxyl radical OH, which is the expected photodissociation product of H$_2$O, was first found in the dayside spectrum of the UHJ WASP-33\,b \citep{nugroho_first_2021}, and confirmed by \citet{cont_atmospheric_2022}. \citet{landman_detection_2021} detect OH in the transmission spectrum of WASP-76\,b, confirmed by \citet{cheverall_robustness_2023}. Retrieval analysis of WASP-33\,b by \citet{finnerty_keck_2023} implies that OH is significantly more abundant than H$_2$O, suggesting the latter is almost completely photodissociated and showing how general assumptions on chemical and physical atmospheric processes can strongly bias retrieved C/O ratios. HCN is detected in the transmission spectrum of HD 189733\,b \citep{cabot_robustness_2019} and HD 209458\,b \citep{hawker_evidence_2018,giacobbe_five_2021}, and tentatively in WASP-76\,b \citep{sanchez-lopez_searching_2022}. 

Metal hydrides and oxides are very prominent in the spectra of late M-dwarfs and Brown Dwarfs, and could therefore also expected to be present in the atmospheres of irradiated gas giants. E.g. Titanium-oxide (TiO) and Vanadium-oxide (VO) have been proposed as high-altitude absorbers to cause thermal inversion layers \citep{2008ApJ...678.1419F}. However, detection of such molecules has been problematic due to sub-optimal line lists \citep{hoeijmakers_search_2015, de_regt_quantitative_2022}, and possibly because they are cold-trapped on the nightside of hot Jupiters\citep[e.g.][]{2013A&A...558A..91P}. The quest for TiO in the dayside spectrum of WASP-33\,b serves as an example of how elusive observational evidence for hydrides and oxides have been. TiO was first detected by \citet{nugroho_high-resolution_2017}. An attempt to confirm this detection with an improved line list using the same data set provided no conclusive evidence \citep{serindag_is_2021}, neither did \citet{herman_search_2020}.  \citet{cont_detection_2021} provide some evidence for TiO, while \citet{yang_high-resolution_2024} report a non-detection. Reports of detections and non-detections seem to come and go, although \citet{prinoth_time-resolved_2023} claim a 6.6 $\sigma$ TiO detection in the transmission spectrum of WASP189\,b.  \citet{pelletier_vanadium_2023} present an exciting detection of VO, while \citet{2023ApJ...953L..19F} find evidence for CrH. The most extensive search for a metal hydride, FeH, involved transmission spectra of twelve planets by \citet{kesseli_search_2020}, finding no signals, corresponding to abundance upper limits of  $<$10$^{-7}$ for most planets. As FeH is a powerful surface-gravity indicator, it is not seen in the spectra of red giant stars. Since transmission spectroscopy probes low atmospheric pressures like those for the photospheres of red giants, it may only be present deeper in the atmosphere.

Species CH$_4$, NH$_3$, and C$_2$H$_2$, have so far only been claimed to be detected using a 4m-class telescope \citep{basilicata_gaps_2024, guilluy_exoplanet_2019,giacobbe_five_2021}. However, such detections have been met with skeptisism, because the combined detections of these molecules are difficult to explain chemically and some are expected to be significantly less abundant than CO and H$_2$O in hot Jupiter atmospheres. Indeed some of these measurements have recently been refuted \citep[][]{2024arXiv240813536B}.

\subsection{Atmospheric temperature structure}

\begin{figure}[t]
\includegraphics[width=5in]{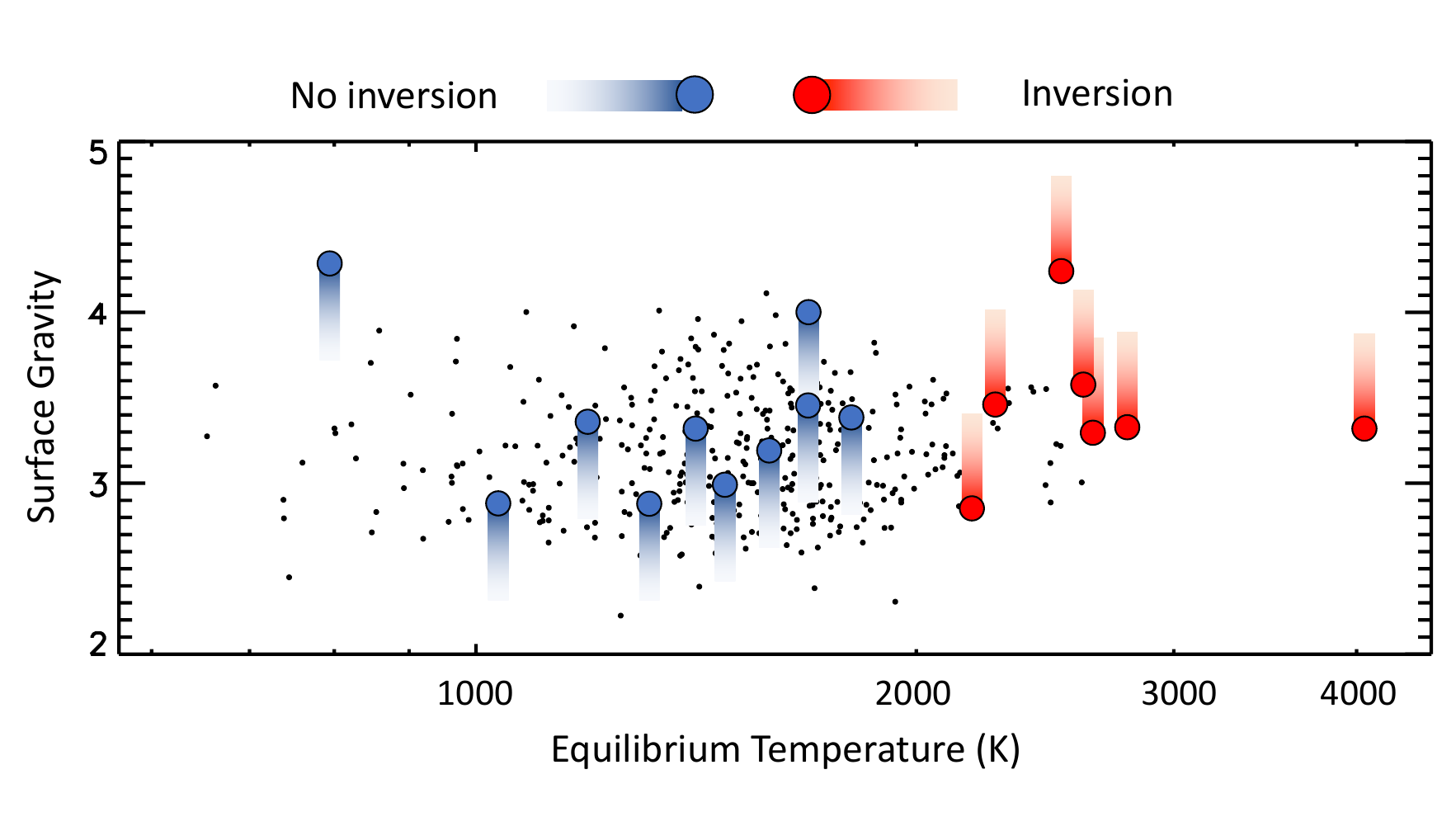}
\caption{Exoplanets with atomic or molecular detections in their dayside spectra, showing either absorption lines indicative of non-inverted atmospheres (blue symbols), or emission lines implying thermal inversions (red). A clear dichotomy is visible at T$_{\rm{eq}} \approx 2000$ K. For non-transiting planets a  radius of 1.25 R$_{\rm{Jup}}$ was assumed. The black dots indicate the general population of transiting gas giants as in Figure \ref{sodium_fig}. References: \citet{pino_neutral_2020};
\citet{guilluy_exoplanet_2019};
\citet{piskorz_ground-_2018};
\citet{bazinet_sub-solar_2024};
\citet{brogi_roasting_2023};
\citet{ramkumar_high-resolution_2023};
\citet{brogi_signature_2012,brogi_detection_2013,brogi_carbon_2014};
\citet{brogi_retrieving_2019};
\citet{de_kok_detection_2013};
\citet{yan_temperature_2020,yan_detection_2022,yan_detection_2022-1,yan_crires_2023};
\citet{birkby_detection_2013,birkby_discovery_2017};
\citet{line_solar_2021}}
\label{inversions}
\end{figure}

HRS phase curve spectroscopy is a very powerful tool to constrain the atmospheric temperature structure of extrasolar planets. Theoretical considerations and 
early spectroscopy with the Spitzer Space Telescope \citep[e.g.][]{2007ApJ...668L.171B,2008ApJ...678.1419F,2008ApJ...673..526K} initiated a discussion about the possible prevalence of thermal inversion layers within the photospheres of hot Jupiters. Such inversions are present in most solar-system planets (albeit not all in their photospheres). The power of high-resolution spectroscopy comes from the fact that in the case of a thermal inversion individual lines are seen in emission, which turned out not to be the case for most hot Jupiters \citep[e.g.][]{2015A&A...576A.111S}. 

Figure \ref{inversions} shows the current results from HRS dayside spectroscopy. Seventeen hot Jupiters and ultra-hot Jupiters have detections from H$_2$O, CO, and/or iron. All observed hot Jupiters show absorption lines and have non-inverted photospheres, while all observed Ultra-Hot Jupiters (T$_{\rm{eq}}$$>$2000 K) exhibit emission lines and have thermal inversions. For a thermal inversion to be present, stellar radiation needs to be absorbed at high altitude. This is most likely caused by species that are efficient absorbers at wavelengths where the host star radiates most of its energy, i.e. the uv/optical wavelength regime. While metal-oxides, in particular TiO and VO, have been proposed to play this role, they appear not to be prominent absorbers in Hot Jupiters and Ultra-Hot Jupiters. Although Sodium absorbs in the optical, it is not present specifically at T$_{\rm{eq}}$$>$2000 K. It has therefore been proposed that iron and other metals, which characteristically are spectroscopically active in the photospheres of UHJs,  may be driving the inversions \citep[e.g.][]{2019ApJ...876...69L,yan_temperature_2020}.

\begin{figure}[t]
\includegraphics[width=4in]{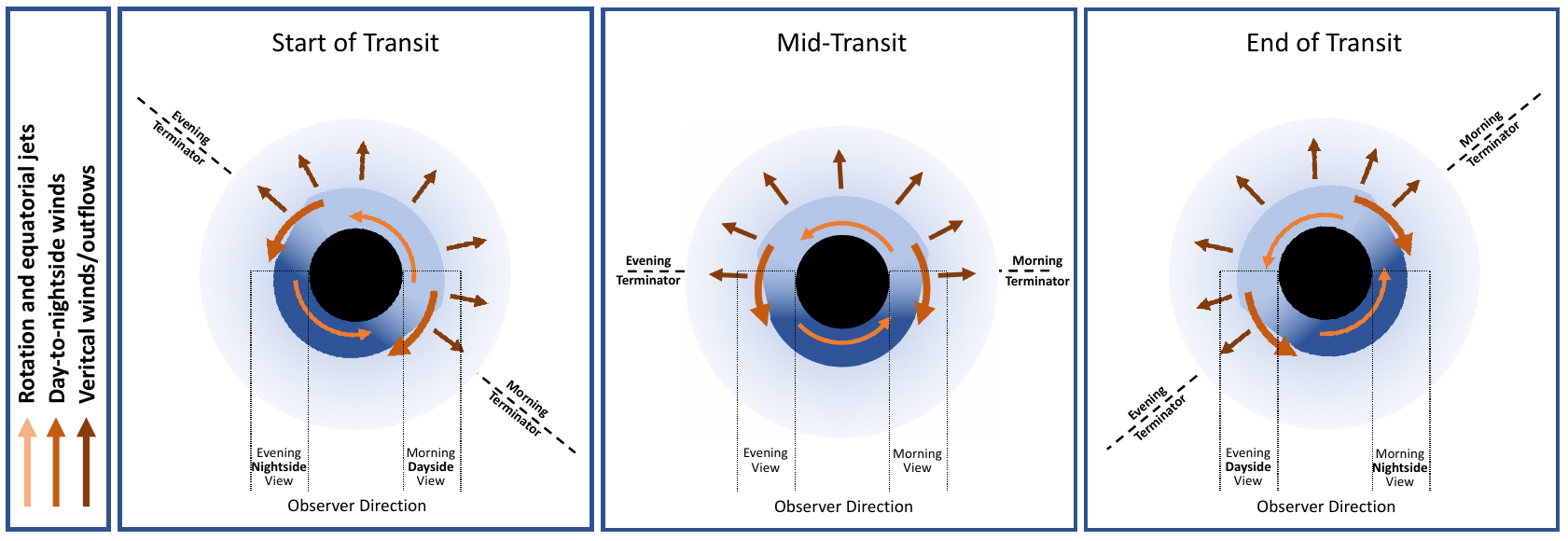}
\caption{Schematic representation of different dynamical effects probed by transmission HRS, with the planet moving from left to right over the stellar surface. these include planet rotation and equatorial jets, day-to-nightside winds, and outflows leading to atmospheric loss. The viewing angle on the terminator region changes during transit. In the first part of the transit the nightside and dayside of the evening and morning limb are probed respectively, and vice versa at the end of the transit.}
\label{dynamics}
\end{figure}
\subsection{Atmospheric dynamics, rotation, and outflows}

\cite{2001ApJ...553.1006B} already advocated more than two decades ago that HRS would be a powerful tool to study atmospheric dynamics. It is indeed an area of research that is unique to this type of observation. The short-period planets that are probed today through transmission and phase curve spectroscopy are expected to be tidally locked, implying that they should have equatorial spin velocities in the range of 1$-$10 km sec$^{-1}$, which is accessable through line-broadening \citep[e.g.][]{brogi_rotation_2016}. In addition, the planetary rotation and strong stellar irradiation are expected to drive complex atmospheric dynamics. 3-Dimensional global circulation models, also including effects like magnetic drag, are developed to capture such effects and link them to state-of-the-art HRS observations \citep[e.g.][]{2008ApJ...682..559S,2010ApJ...714.1334R, showman_doppler_2013, 2013A&A...558A..91P, kempton_high_2014,2022AJ....163...35B}. Fig. \ref{dynamics} illustrates the different effects to which HRS is potentially sensitive. Planetary rotation with strong stellar forcing is expected to induce equatorial jets that amplify line broadening. Global day-to-nightside winds 
may transport heat from the dayside to the nightside, which will blueshift transmission lines. In addition, strong stellar irradiation may induce vertical Parker-type winds leading to outflows and atmospheric loss.  For the most extreme close-in planets (in terms of a/R$_*$) the orientation of the planet changes significantly during transit, meaning that different parts of the evening and morning limbs can be studied, resulting in fascinating effects \citep{ehrenreich_nightside_2020}.  
 
 \begin{figure}[t]
\includegraphics[width=4.5in]{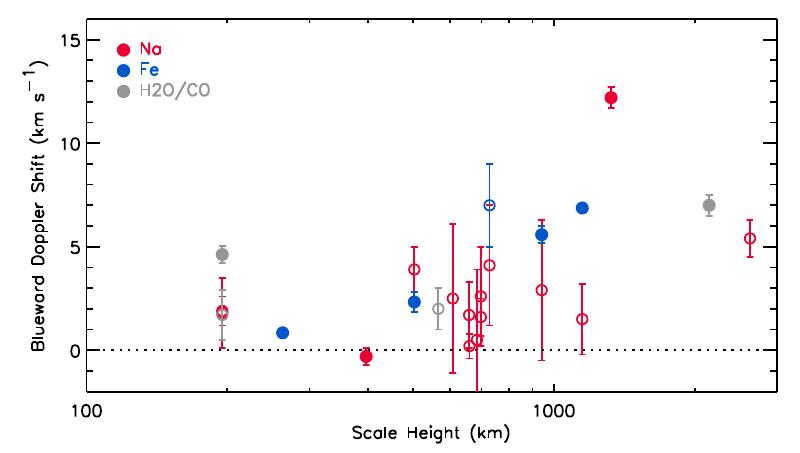}
\caption{Day-to-nightside winds, as derived from Doppler shifts in transmission spectra, as function of atmospheric scale height. Data points in blue, green and grey are based on Fe, Na, and CO/H$_2$O respectively,  and open circles if their uncertainties are $>1$ km s$^{-1}$. Although this is a very heterogeneous and incomplete sample, those planets with larger scale heights appear to have, on average, larger day-to-nightside winds. Data from: \citet{Langeveld_survey_2022,gandhi_retrieval_2023,pai_asnodkar_variable_2022,boucher_characterizing_2021,boucher_co_2023,louden_spatially_2015,brogi_rotation_2016,snellen_orbital_2010,mounzer_hot_2022,seidel_detection_2023,chen_detection_2020,wyttenbach_hot_2017,rahmati_transmission_2022}
}
\label{DayNight}
\end{figure}

\paragraph*{Day-to-nightside winds}

Winds that travel from the dayside to the nightside result in a global blueshift of transmission lines. Their velocities are thought to result from a balance between the day/night temperature difference and dissipation mechanisms \citep[e.g.][]{miller-ricci_kempton_constraining_2012}. First evidence of such winds, although at low significance, was presented by \citet{snellen_orbital_2010} using CO transmission lines in HD 209458\,b. Before conclusions can be drawn from Doppler shift measurements of transmission lines, it must be realised that several effects can influence such observations. The orbital ephemeris needs to be well constrained. For a planet in a 2-day orbit around a solar-type star, a 1 minute error in transit-timing results in a 0.4 km s$^{-1}$ error in the expected radial velocity of the planet, while an error in eccentricity of 0.01 can change the planet's expected velocity during transit by 1 km s$^{-1}$. Also, the system velocity can be difficult to determine for a fast-rotating star. The stellar gravitational redshift  (0.6 km s$^{-1}$ for our Sun) needs to be taken into account, as well as the possible blueshift due to convection cells. If elemental transitions are used that are also present in the stellar atmosphere, such as sodium or iron, an imperfect correction for the Doppler shadow and/or CLV effects may also affect wind measurements.

Reports of day-to-nightside winds have been rather scattered so far, mostly focusing on a single elemental species or molecule for a single planet. Uncertainties are difficult to assess, with different studies of the same planet, even using the same species, sometimes resulting in significantly different wind speeds. \cite{pai_asnodkar_variable_2022} show that even variability may play a role, with four measurements of iron in the transmission spectrum of KELT-9\,b taken at different epochs resulting in blueshifts ranging from 4 to 11 km s$^{-1}$. There have only been a few systematic studies of wind effects in exoplanet atmospheres. Notably, \cite{Langeveld_survey_2022} studied ten hot Jupiters and ultra-hot Jupiters in sodium to find blueshifts up to 5 km s$^{-1}$, but all-bar-one within 2$\sigma$ from zero $-$ hence barely significant. The retrieval study by \cite{gandhi_retrieval_2023} of six ultra-hot Jupiters show them all to have a net blueshift, from $<$1 km s$^{-1}$ for MASCARA-2\,b to 7 km  s$^{-1}$ for WASP-76\,b, mostly at higher statistical significances.  

Figure \ref{DayNight} is an attempt to reveal a pattern in the very heterogeneous and incomplete data on atmospheric winds, based on different atmospheric species. Warm and fluffy planets possibly exhibit stronger winds: all planets with measured winds $>$ 5 km s$^{-1}$ have atmospheric scale heights $>$700 km. 

\paragraph*{Rotation, equatorial jets and time-resolved spectroscopy}

While day-to-nightside winds are expected to dominate at low pressures high-up in hot-Jupiter atmospheres, most heat transport from the day to nightside is predicted to come from equatorial jets deeper in the atmosphere. Indeed, space-based thermal phase-curve observations of hot Jupiters often show \citep[e.g.][]{2007Natur.447..183K} that their hottest parts are significantly offset from a planet's substellar point.  HRS observations can directly probe these equatorial jets through line broadening in addition to what is expected from bulk rotation of a tidally locked planet \citep[e.g][]{louden_spatially_2015,brogi_rotation_2016}. However, different contributions from the leading and trailing limbs and contributions from the day-to-nightside winds, thermal broadening, and possible residuals from the RM effect, can significantly complicate such analyses. It is therefore very helpful to add extra information using time-resolved transit spectroscopy. Contributions from the leading (morning side) and trailing (evening side) limb can be isolated by focusing on the ingress and egress transmission spectrum respectively. In this way, \cite{louden_spatially_2015} derive wind velocities on the leading and trailing limb of HD 189733\,b that are consistent with predictions from atmospheric circulation models and can be understood as a combination of tidally locked planetary rotation and an eastward equatorial jet. \cite{seidel_wind_2020, seidel_into_2021} retrieved wind patterns, speeds, and temperature profiles from the line shape of the sodium doublet of the ultra-hot Jupiter WASP-76\,b, showing that this is best fit with a  uniform day-to-nightside wind and a significant vertical wind (see below). More recently, they used the unprecedented collecting area of the four VLT telescopes combined with ESPRESSO's 4UT mode to probe the sodium line in the ultra-hot Jupiter WASP-121\,b \citep{seidel_detection_2023-1} and identify a high-velocity component during egress interpreted as an equatorial day-to-night side wind across the evening limb. 
 
\cite{ehrenreich_nightside_2020}  realised that for planets in particularly close-in orbits, i.e. with small a/R$_*$ (ratio of orbital distance to stellar radius), also the orientation of the planet changes significantly during transit. Due to this {\sl Ehrenreich Effect}, the viewing angle of WASP-76\,b varies by about 30$^\circ$ from ingress to egress (see Figure \ref{dynamics}). It means that early on in the transit, the transmission spectrum probes on the morning limb the atmospheric dayside of the planet, and on the evening limb the nightside. Late in the transit this is reversed and the dayside evening limb and nightside morning limb are probed. Since contributions from both limbs can be separated because of their different Doppler shifts (due to planet rotation), the planet atmosphere can be traced from day to night-side on the evening and morning side. For WASP-76\,b \cite{ehrenreich_nightside_2020} show that this results in a transmission signal in iron that increases significantly in strength and blueshift during transit \citep[observation confirmed by][]{kesseli_confirmation_2021}, interpreted  as due to a lack of  iron on the morning-nightside due to rain-out. \cite{wardenier_decomposing_2021} show using 3D Monte Carlo radiative transfer modelling that the signal could also be caused by a substantial temperature asymmetry. \cite{gandhi_spatially_2022} conducted an atmospheric retrieval analysis  separately for the morning and evening terminator regions  finding evidence for both a lower iron abundance and lower temperature at the morning limb.

While WASP 76\,b is the optimal target to probe the Ehrenreich Effect (among all studied Ultra-Hot Jupiters), and iron the optimal species, similar effects have been detected for other metals and other planets \citep[e.g.][]{bourrier_hot_2020, stangret_detection_2020, hoeijmakers_high-resolution_2020, kesseli_atomic_2022}. 

In addition to time-resolved transmission spectroscopy, emission spectroscopy can also provide information about heat circulation in combination with global circulation models \citep[e.g.][]{2022A&A...668A.176P,van_sluijs_carbon_2023}.

\paragraph*{Outflows and atmospheric loss}

In the years after the first detection of an exoplanet atmosphere \citep{2002ApJ...568..377C}, the archetypical transiting planets HD 209458\,b and HD 189733\,b were shown 
to exhibit Lyman alpha absorption at levels of a few percent during transit \citep{2003Natur.422..143V,2010A&A...514A..72L}, caused by a significant halo of exospheric hydrogen. The warm Neptune GJ 436\,b was found to have such halo extending as a cometary tail around the planet causing a 50\% transit depth \citep{2015Natur.522..459E}. Unfortunately, probing hydrogen through Lyman alpha is challenging due to geocoronal absorption which blocks all potential signals near zero velocity, and absorption by the interstellar medium beyond the Local Bubble (30 - 100 pc).  Transmission signals by hydrogen from higher energy levels, such as H-alpha \citep{yan_extended_2018}, other Balmer lines \citep{wyttenbach_mass-loss_2020,yan_detection_2021}, and even Paschen-beta \citep{sanchez-lopez_detection_2022} are accessible with HRS and point to exospheric gas associated to atmospheric escape, but seem restricted to  planets with the highest temperatures \citep{zhang_transmission_2022}.  

The seminal work by \cite{2000ApJ...537..916S}  predicted the presence of helium absorption at 1083 nm in the transmission spectra of hot Jupiters. After some failed attempts to detect this transition, it was overlooked for almost two decades until interest was rejuvenated by \cite{2018ApJ...855L..11O} and subsequently observed with HST \citep{2018Natur.557...68S} as a direct probe of atmospheric loss. Since it is in a wavelength region relatively free of telluric absorption, ground-based HRS has proven to be a very powerful technique to probe its spatial and dynamical structure \citep{2018Sci...362.1384A,nortmann_ground-based_2018}. This has revolutionised the study of atmospheric escape, although interpretation of the 1083 nm helium line and conversion to atmospheric losses is less straightforward than that for Lyman-alpha. The helium line originates from an excited metastable 2$^3$S state for which the population is governed by the radiation field of the host star: while extreme-ultraviolet flux ionises the helium ground state, mid-ultraviolet flux ionises the metastable state \citep{2019ApJ...881..133O}.  Close-in planets around K-stars are expected to show the strongest helium absorption while those orbiting early-type stars, despite probably having high mass-loss rates, may show no signal since the metastable state is not sufficiently populated \citep[e.g KELT-9\,b;][]{nortmann_ground-based_2018}. 

About a dozen planets have been detected in He I, some at very high signal to noise. Both the warm Saturns WASP-69\,b and WASP-107\,b, located near the edge of the evaporation desert, are shown to have strong absorption signatures with blueshifts of several km s$^{-1}$ and post-transit absorption pointing to comet-like tails.  \citep{nortmann_ground-based_2018,2019A&A...623A..58A}. For the latter it is estimated that its extended thermosphere fills half of the Roche lobe with an estimated escape rate of meta-stable helium of 10$^{6}$ g s$^{-1}$ \citep{2019A&A...623A..58A}. Recently, the cometary tail of WASP-69\,b was found to extend to 7.5 planetary radii behind the planet \citep{2024ApJ...960..123T}. Helium absorption signals can be significantly more extended and complex than this. HAT-P-67\,b is exhibiting helium absorption in a large {\sl leading} trail up to 130 planetary radii \citep[necessarily observed over multiple nights;][]{2024AJ....167..142G}. Giant tidal tails of escaping helium have been found both leading and trailing the hot Jupiter HAT-P-32\,b \citep{2023SciA....9F8736Z} spanning a projected length over 53 planetary radii. It is the intricate interplay between orbital shear, day/nightside mass-loss asymmetries and confinement by the stellar wind that is expected to determine the spatial and dynamical morphology of exospheric helium \citep{2019ApJ...873...89M}. By the absence of any stellar wind, atmospheric loss from the planet dayside and nightside will cause a leading and trailing tail respectively due to orbital shear. A strong stellar wind will push any material into a cometary tail behind the planet. \cite{zhang_search_2020} suggest that the 1083 nm helium line may also be observed as airglow emission away from transit, but that this may require one of the upcoming extremely large telescopes. Helium studies have also been extended to mini-Neptunes \citep[e.g.][]{2023AJ....165...62Z}.

\subsection{Reflected-light searches}

The field of exoplanet atmospheric characterisation once started with attempts to use ground-based HRS observations to probe reflected light, i.e. starlight scattered by planet atmospheres  \citep[e.g.]{1999ApJ...522L.145C,1999Natur.402..751C}.  Unfortunately, progress in this particular avenue has been very difficult ever since. While scattered light signals could in principle be as high as $\sim$10$^{-4}$ for highly reflective (i.e. cloudy) atmospheres, Hot Jupiters turn out to have very low geometric albedos. Many studies have focused on the tau Bootis system \citep{1999ApJ...522L.145C,2003MNRAS.344.1271L,hoeijmakers_searching_2018} because both star and planet are tidally locked. This means that the star is stationary in the planet rest frame, cancelling out the velocity broadening in the reflected light spectrum and maximising a potential atmospheric signal \citep[see also][]{spring_black_2022}. 

\cite{2015A&A...576A.134M} provided evidence for a reflected light signal from 51 Peg\,b, pointing to a planet radius of 1.9 R$_{\rm{{Jup}}}$ and a high geometric albedo of 0.5. However, subsequent studies have not been able to confirm the signal \citep{spring_black_2022}. It is well possible that HRS reflected light detections will need to wait for the future ELTs. In particular the combination of HRS with high-contrast imaging probing neighbouring stars will be orders of magnitudes more sensitive to scattered light signals, and will be able to probe a new regime of cooler planets that are expected to have significantly higher geometric albedos than Hot Jupiters \citep{lovis_atmospheric_2017}. 

\subsection{Chromatic Doppler Shadow measurements}

Originally proposed two decades ago \citep[][]{snellen_new_2004,dreizler_possibility_2009}, the RM effect or Doppler shadow can in principle also be used to probe exoplanet atmospheres. At wavelengths where the effective planet size (due to extra atmospheric absorption) is larger, the amplitude of the RM effect and Doppler shadow will also be larger. In measuring the amplitude of the RM effect as function of wavelength a transmission spectrum can be constructed. In this way, \cite{di_gloria_using_2015} reconstructed the broadband transmission spectrum of HD 189733\,b showing evidence, albeit at low significance, for a Rayleigh scattering slope similar to that found using HST \citep{2011MNRAS.416.1443S}. Independent observations and analyses come to comparable conclusions \citep{cristo_carm_2022}.
For HD 209458\,b, a remarkable similarity was found between the broadband RM-transit spectrum and that from HST by \cite{2020A&A...644A..51S}. \cite{2019A&A...631A..34B,2021A&A...645A..24B} found chromatic RM effects for ultra-hot Jupiters, but do not present  transmission spectra. 

Analysis of the chromatic Doppler shadow remains highly challenging.  While based on shape deformations of stellar lines, these same lines may also be present in the planet transmission spectrum (as in the case of sodium, or any metal lines in the case of UHJs), highly complicating the analysis. Depending on the orbital configuration (i.e. spin-orbital alignment) and intrinsic properties of a planet's atmospheric absorption, the strength of the Doppler shadow may significantly vary during transit. This is most apparent when the planetary absorption and Doppler shadow overlap in velocity. These specific epochs are often left out of transmission spectroscopic analyses. However, with the limited lifetime of HST, the chromatic Doppler shadow or RM effect may at some point be the only way to obtain information on blue-optical broadband transmission spectra in the near future. 

\section{Results on wide-separation exoplanets}
 
Compared to HRS transit and phase curve measurements, HRS combined with high-contrast imaging is still in its relative infancy. First observations of this type were performed at medium spectral resolution, such as the important work by \cite{2013Sci...339.1398K}, while \cite{snellen_fast_2014} showed its potential by measuring the radial orbital velocity and $v\sin{i}$ of the Super-Jupiter $\beta$ Pictoris\,b. In this section I will highlight some of the state-of-the-art observations, focusing on the first results from KPIC on Keck, and CRIRES+.  

The larger the angular distance between planet and host star, the more accurately the planet spectrum can be separated from that of the star. In many cases, the high-resolution spectra that are obtained with this technique are of significantly better quality than transmission or phase-curve spectra. In this section,  some recent results are also included from isolated brown dwarfs for which the spectra reach signal-to-noise ratios of $>$50 per spectral resolution element. In addition to giving potential insights into different formation pathways for brown dwarfs and planets, such observations serve as an ideal benchmark to test the accuracy and robustness of analysis techniques, and to gauge the level of detail that can be retrieved in the case of future high-quality exoplanet spectra.

\subsection{ Molecular abundances and atmospheric temperature structure}
 
In the last 2-3 years, HRS studies of directly imaged planets and brown dwarfs have not only resulted in detections of many molecular species, but also in stringent constraints on abundances and temperature-pressure profiles. About two-dozen substellar objects have now been targeted with HRS at high signal-to-noise allowing reliable retrieval of atmospheric parameters. We are at the start of a wave of many such investigations which so far have mainly targeted atmospheres in K-band where H$_2$O and CO are the two dominant contributors for hot/warm objects $>$1500 K, with an increasing contribution from CH$_4$ and NH$_3$ towards lower temperatures \citep[e.g.][]{wang_detection_2021,2022ApJ...937...54X,Costes2024,deregt2024}. The hotter targets also show clear absorption by hydrogen fluoride (HF) and several atomic species such as sodium, calcium, and titanium \citep{2024arXiv240707678G}. 

\begin{figure}[h]
\includegraphics[width=4.5in]{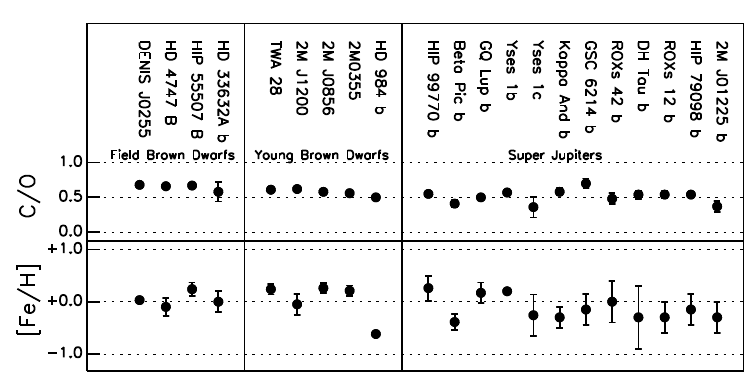}
\caption{HRS atmospheric retrieval results on C/O ratio and metallicity (derived from C/H) as present in the literature, for field brown dwarfs, young brown dwarfs, and Super-Jupiters. For some measurements the error bars are smaller than the symbol sizes. Data and references are provided in Supplementary Table 2.}
\label{metallicity}
\end{figure}

Figure \ref{metallicity} shows an overview of a heterogeneous sample of young and older field brown dwarfs and super Jupiters for which metallicities and C/O ratios have been measured. Since these are significantly more massive than Jupiter (10 - 50 M$_{\rm{Jup}}$) it may not be surprising that their metallicities are broadly consistent with that of our Sun, except for Beta Pictoris b and HD 984 b, which appear to have a lower metal content \citep[][]{landman_betapic_2024,Costes2024}.  The metallicities of the Super-Jupiters seem on average somewhat lower than those of the brown dwarfs, but this is unlikely to be statistically significant. They have generally lower-quality spectra resulting in larger error bars and therefore may be biased. Also, the well known log(g)-metallicity correlation adds to the uncertainty. While C/O ratios could potentially provide insights in their formation processes, almost all objects are consistently measured to be within the C/O=$0.5 - 0.65$ range. 

As a detailed case study, the work by \cite{deregt2024} is discussed, who obtained a K-band spectrum of the T=1400 K field brown dwarf DENIS J0255-4700 at a signal-to-noise of $40$ per resolution element as part of the CRIRES+ SupJup Survey. They compared atmospheric retrievals using three different set ups: 1) Free chemistry in which the abundances of relevant chemicals are constant with pressure but otherwise unconstrained; 2) Equilibrium chemistry in which the chemical abundances at each pressure are set by the metallicity, C/O ratio and temperature; and 3) Quenched equilibrium chemistry in which the abundance follow chemical equilibrium down to a certain pressure, below which the abundances are forced to be constant. The physical idea is that below this quench pressure (so, higher up in the atmosphere) the vertical mixing time scales are shorter than the chemical time scales such that the chemical abundances are governed by the circumstances at the quench pressure. Interestingly, the high data quality allows distinctions to be made between these scenarios. The free chemistry and quench-pressure set-up result in a very similar solution in temperature/pressure profile, chemical abundances, and C/O ratio. The un-quenched chemical equilibrium model provides a significantly worse fit, resulting in a different metallicity and C/O ratio. Basically, the relatively small methane signal pushes the chemical equilibrium model to higher temperatures. At lower signal-to-noise, one may have chosen a model with the lowest number of parameters (i.e. following the Bayesian Information Criterion), which is the chemical equilibrium model, possibly skewing the retrieved parameters. In more extreme cases, the lack of methane, while it is expected at the derived atmospheric temperatures, points to quenching \citep[e.g.][]{2024AJ....168..131Z}. 

Atmospheric retrieval set-ups also allow for the introduction of clouds, either grey or physically motivated (e.g. MgSiO$_3$, Fe, etc.). However, so far, most retrieval analyses find little evidence for significant cloud opacity. Finding evidence for clouds is generally challenging for HRS observations over a relatively narrow wavelength range, where slight changes in the temperature structure can mimic clouds. Most analyses have so far been performed in K-band which probes relatively high altitudes and clouds may be present below the photosphere \citep{2022ApJ...937...54X}. J and H-band HRS spectra should be significantly more sensitive to cloud layers. As  shown by \cite{deregt2024}, the K-band spectrum of DENIS J0255 provides no statistical evidence for cloud opacity, but the cloud-free model-flux at J-band is a factor $\sim$3 higher than measured by 2MASS photometry, meaning that clouds indeed need to be present but possibly below the K-band photosphere.

High signal-to-noise spectra also allow detailed modelling of the temperature structure, e.g. by parametrising the T/p profile with temperature gradients \citep{2023AJ....166..198Z} to retrieve the radiative-convective boundary \citep{2015ARA&A..53..279M}, as is done for GQ Lupi b (Gonzalez Picos et al. 2024, submitted). In addition, the retrieval of a slight deviation from the adiabatic temperature gradient at high pressure suggests the presence of non-adiabatic convection or the cloud-P/T degeneracy \citep[e.g.][]{2019ApJ...876..144T}. An additional effect that has been identified in high signal-to-noise spectra is veiling \citep{1999A&A...352..517F}. Some young substellar objects may have a circum-planetary disk that significantly contributes to the observed spectrum as a continuum source, making the object spectral lines appear shallower. E.g. significant veiling at $\sim$14\% of the continuum flux at K-band is measured for the young brown dwarf 2M J1200 \citep{2024arXiv240707678G}, and it is found to be very prominent in the K-band spectrum of the GQ Lupi host star at $>$50\% level (Gonzalez Picos et al 2024, submitted). 
 
The higher the data quality, the more complex the modelling needs to be to appropriately fit the data. While K-band HRS monitoring of the nearby brown dwarf Luhman 16\,B  resulted in the first brown-dwarf Doppler-imaging map \citep{crossfield_doppler_2014}, De Regt et al. (in preparation) show that at S/n$\ge$100/pixel, HRS becomes sensitive to spatial inhomogeneities. This can reveal the presence of cloud bands and vortices in single-epoch observations by deviation from the expected rotational line-broadening profile from a homogeneous limb-darkened disk. Other model refinements are possibly required at the highest levels of data quality, such as molecular abundance variations with altitude due to non-LTE processes.
  
\subsection{Orbital and Rotational Dynamics}

The atmospheric retrieval analyses as discussed above also provide the projected rotational broadening, or $v\sin{i}$, of the observed targets. While $\sin{i}$ is unknown for most objects, for a randomly oriented sample,  the true spin velocity is on average only 15\% higher. Also, only 15\% of a sample of randomly oriented targets is expected to have i$<$30$^\circ$ with true spin velocities that are larger by a factor two or more than the measured $v\sin{i}$.  In principle, their inclinations can be determined if they show significant flux variability that lead to the derivation of their spin period. In combination with constraints on their orbital inclination, the radial component of the obliquity can be determined, as shown by \cite{2020AJ....159..181B} who derive a planetary obliquity for the Super-Jupiter 2M J0122 b of at least 49$^{+27}_{-21}$ degrees.

\begin{figure}[h]
\includegraphics[width=4in]{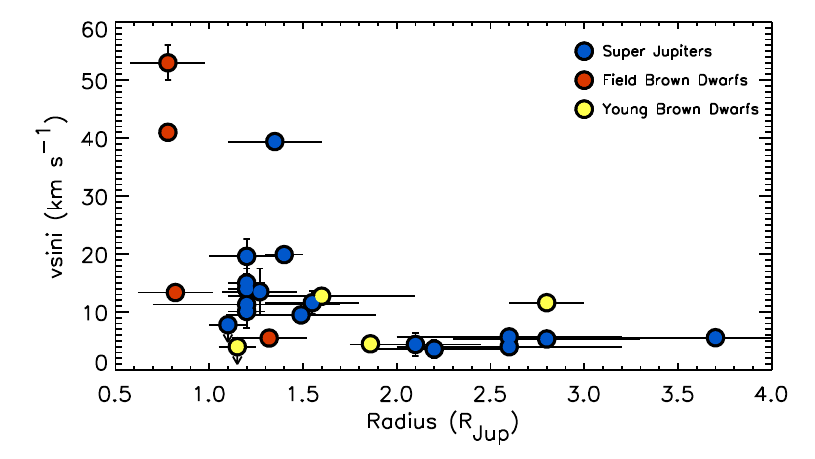}
\caption{HRS measurements of v$sin$i as function of estimated object radius for field brown dwarfs (red), young brown dwarfs (yellow) and Super-Jupiters (blue). Data and references are provided in Supplementary Table 2.}
\label{spin}
\end{figure}

Investigating the rotation rates of exoplanets may shed light on their formation mechanisms, how their angular moment is set via interaction with a circum-planetary disk through magnetic coupling,  and possibly altered later through collisions or gravitational tides \citep{bryan_constraints_2018,bryan_as_2020}. For those substellar objects with measured $v\sin{i}$  the radii can be estimated from their derived surface brightness, flux density, and distance to Earth. Figure \ref{spin} shows the measured $v\sin{i}$ as function of radius for field brown dwarfs, young brown dwarfs, and Super-Jupiters. In any evolutionary sequence it would be expected that the youngest, bloated objects rotate the slowest, subsequently cool and contract over time and spin up. Although this is a very heterogeneous sample, all-bar-one objects with R$>$1.75 R$_{\rm{Jup}}$ have a $v\sin{i}<$10 km s$^{-1}$, while most targets with R$<$1.75 R$_{\rm{Jup}}$ exhibit significantly higher spin velocities. In the future, large and complete samples of brown dwarfs and gas-giant exoplanets for which the ages, radii, and masses can be derived more accurately, will provide interesting constraints on the evolution of angular momentum of these objects. While the current sample is a varied assortment, and the uncertainties in radii are large, the distribution of $v\sin{i}$ for Jupiter-size objects seems to have a larger spread than expected from a random distribution in $\sin{i}$ alone. Objects of the same size seem to have significantly different rotation velocities. 

In addition to rotational broadening, atmospheric retrievals also reveal the radial velocity of the exoplanet, and if that of the host star is determined, the radial component of its orbital velocity. This can be used to constrain the planet orbital elements, as shown by \cite{schwarz_slow_2016} for GQ Lup\,b. There is growing evidence that some Super-Jupiters at wide orbits have high eccentricities that point to a formation in a close-in orbit and subsequent planet-planet interactions \citep[e.g.][]{2023MNRAS.519.1688D}. 
  
\subsection{Isotope measurements}

\begin{figure}[h]
\includegraphics[width=4.5in]{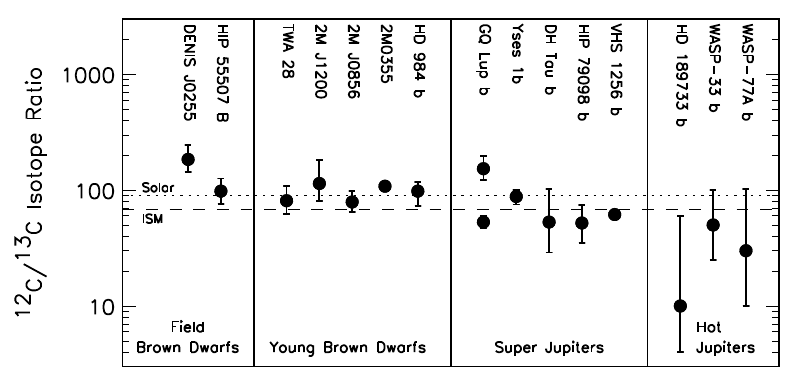}
\caption{Carbon isotope ratios presented in the literature for field brown dwarfs, young brown dwarfs, Super-Jupiters, and Hot Jupiters. Data and references are provided in Supplementary Table 2.}
\label{isotopes}
\end{figure}

The ability to detect individual isotopologues and determine isotope ratios in exoplanet atmospheres opens up a new pathway to constrain planet formation and evolutionary processes \citep[e.g.][]{molliere_detecting_2019,2019ApJ...882L..29M}. Figure \ref{isotopes} shows the derived $^{12}$C/$^{13}$C isotope ratios from HRS retrieval analysis for brown dwarfs, Super-Jupiters and hot Jupiters. Again, this is a very heterogeneous sample and could be biased in multiple ways. The $^{12}$C/$^{13}$C ratio for VHS1256\,b is derived from medium-resolution JWST spectroscopy, which also detects $^{17}$O and $^{18}$O \citep{gandhi_jwst_2023}. The dashed and dotted lines indicate the average isotope ratio in the interstellar medium and the solar system respectively. The first exoplanet isotope measured as presented by \cite{zhang_13co-rich_2021} showed the atmosphere of YSES 1\,b to be rich in carbon-13 with $^{12}$C/$^{13}$C = 31$^{+17}_{-10}$, but higher-quality HRS observations now point to $^{12}$C/$^{13}$C = 88$\pm$13 \citep{2024arXiv240916660Z}. Also, two independent HRS studies of GQ Lup \,b point to significantly different isotope ratios. It implies that these measurements are still challenging, that uncertainties may be underestimated, and/or that systematic effects are yet not well understood. Taking the measurements in Fig. \ref{isotopes} at face value, the Super-Jupiters appear to have slightly lower carbon isotope ratios than the brown dwarfs, while the three hot Jupiters have even lower ratios. However, these latter measurements are presented as tentative detections and may suffer from detection bias. It is fair to say that it is too early to draw any conclusions, but that a new avenue of exoplanet atmospheric characterisation is opening up.

\section{Synergies with JWST spectroscopy of exoplanets}

Only two years after the start of its science operations, it is clear that the James Webb Space Telescope is revolutionising the field of extrasolar planets. While transmission spectroscopy of Hot Jupiters has revealed previously undetected species such as CO$_2$ and SO$_2$ \citep[e.g.][]{2024Natur.626..979P},  it also pioneers detailed atmospheric characterisation of smaller and cooler planets. For example, SO$_2$ and silicate clouds are detected in a warm Neptune \citep{2024Natur.625...51D}, strong evidence is found for a secondary atmosphere around the hot rocky exoplanet 55 Cancri e \citep{2024Natur.630..609H}, and mid-infrared secondary eclipse observations with MIRI are constraining the dayside temperatures of the Earth-size planets in the TRAPPIST-1 system \citep[e.g.][]{2023Natur.620..746Z}. 

The great sensitivity of JWST relative to ground-based observations, in particular at wavelength beyond 2.5 $\mu$m, means that it will most likely require the enormous collecting area of the ELTs for HRS to play a competitive role in the thermal infrared. We will discuss how HRS will break new ground in the era of the ELTs in the next section. However, there  are exciting synergies with JWST with current 8-10m class telescopes, and domains where ground-based HRS continues to provide uniquely new insights. 

Some spectral regions can be particularly crowded with absorption features from a variety of species such that low resolution spectroscopy cannot unambiguously identify the absorbing molecules or atoms. The optical/near-infrared spectra of Ultra-Hot Jupiters are very rich in metals (see Fig. \ref{chemicals}) which are challenging to individually identify with JWST's limited spectral resolution.  Also the velocity structure of exospheric helium, as a probe of atmospheric loss process, is uniquely accessible with ground-based HRS. The spectral signature of CO, consisting of series of regularly spaced lines, is particularly favourable for HRS. While CO was the first molecule detected in the transmission spectrum of HD 209458\,b \citep{snellen_orbital_2010}, it is barely seen in JWST/NIRSpec transmission observations of the planet \citep{xue_jwst_2024}. 
Also, CO is not seen in JWST/NIRISS secondary eclipse observations of WASP-18\,b \citep{2023Natur.620..292C}, while clearly detected in ground-based HRS observations with IGRINS \citep{brogi_roasting_2023} allowing for the planet's C/O ratio to be measured. 

The study of minor isotopes has  exciting synergetic opportunities between HRS and JWST observations. While ground-based HRS has so far focused on $^{13}$CO, JWST observations have revealed $^{15}$NH$_3$ \citep{2023Natur.624..263B} and oxygen-17 and oxygen-18 isotopes of CO \citep{gandhi_jwst_2023}, and possibly deuterium (in CH$_3$D) is within reach \citep{2019ApJ...882L..29M}.  Future will tell whether trends in isotope ratios will emerge that will shed new light on planet formation and evolutionary processes.  

The spectral resolution of JWST instrumentation is insufficient to directly probe any dynamics, but phase curve analyses of close-in planets still provide crucial information on day-nightside heat distribution, heat transportation, and longitudinal distribution of molecular gases. These can be coupled to direct measurements of wind-speeds and morning/evening terminator retrievals from HRS to build up a holistic picture of hot Jupiter atmospheres. In a similar way, JWST variability monitoring can be directly linked to HRS $v\sin{i}$ and line profile variation measurements to constrain cloud cover distributions and planet obliquities. 

While HRS preserves information on individual line profiles and probes atmospheres at relatively low pressures, JWST observations preserve continuum information more accurately, have generally a wider spectral range and probe deeper parts of the atmosphere. Combining the two can be a powerful way to constrain a planet's properties \citep{brogi_framework_2017,2019AJ....158..228G}. \cite{2024AJ....167..110S} used a joint retrieval framework on secondary-eclipse JWST and pre- and post-eclipse HRS observations on the hot Jupiter WASP-77A\,b, showing that both data sets are consistent and that molecular abundances as well as vertical thermal structure could be determined at higher precision than by using either data set individually. It is also interesting to apply ground-based cross-correlation HRS techniques to medium-resolution JWST observations. \cite{2017AJ....154...77S} proposed to apply such a technique to MIRI/MRS spectra of Proxima Cen\,b to reveal potential signatures from CO$_2$, which seems technically within the realm of possibilities \citep{2024PASP..136h4402D}. \cite{2023ApJ...955L..19E} apply cross-correlation techniques on WASP-39\,b NIRSPEC data showing it to be more powerful than a conventional analysis, unequivocally showing the presence of CO and providing evidence for $^{13}$CO. 

It remains a fact that JWST is pushing exoplanet characterisation to a regime that currently seems out of reach of ground-based HRS. It remains to be seen how far HRS transmission spectroscopy can be pushed to cooler planets on wider orbits. The question is whether ground-based HRS observers should be more bold and daring. It is important to know whether detections of CO$_2$ and SO$_2$ are currently too difficult from the ground, or whether simply nobody tried. 
Maybe HRS observers (and telescope allocation committees) have been too conservative and not willing to invest significantly more observing time on individual targets to break new ground. The same can be argued for the new discoveries of young embedded planets, such as PDS 70  \citep{haffert_two_2019,cugno_molecular_2021} and AB Aur \citep{2022NatAs...6..751C}, for which HRS observations could constrain the dynamics and chemistry of their accretion environment. 

\section{Future opportunities with the ELT}

As discussed in Section 2, with the advent of the ELTs we will enter a new era of HRS exoplanet characterisation. The increase in collecting area (and detection speed) of up to a factor 25 will increase the sensitivity of HRS transmission and phase-curve observations by up to a factor 5. It means that high-resolution emission and transmission spectra of hot gas giant planets will be obtained of similar quality as those currently obtained for brown dwarfs \citep{2024AJ....168..133P}, but that also temporal information, e.g. during ingress, egress or specific orbital phases will provide detailed information of the temperature structure, gaseous content and dynamics in specific areas of their atmospheres.  While transmission spectroscopy of cooler exoplanets often result in flat transmission spectra due to clouds and hazes, it is expected that ELT-HRS can probe such atmospheres above their clouds to determine their gaseous contents and climates \citep[e.g.][]{gandhi_seeing_2020,2020AJ....160..198H,2024A&A...688A.191G}. Also, it may well be that HRS with the ELTs will break the degeneracies originating from stellar activity that currently plague JWST transmission observations \citep[e.g][]{2024AAS...24440705D}. 

It is high-contrast HRS where enormous gains will be made, increasing the detection speed by three orders of magnitude and unlocking exoplanets at unprecedentedly small angular distances from their host stars. While METIS \citep{2021Msngr.182...22B}, first-generation instrument at the ELT, will have R=100,000 integral field and long-slit spectroscopy between 3 and 5 $\mu$m at near-maximum Strehl ratios, second-generation instrument ANDES (which will exhibit a small IFU) will serve the optical and near-infrared regime at similar spectral resolution \citep{2022SPIE12184E..24M}.  While current instrumentation is generally limited to such contrasts that it can {\sl only} probe young planets that are still sufficiently warm from their formation, the ELTs are expected to push into the regime that scattered light from planets around our nearest neighbours can be studied.  Many such planets are already known to exist through radial velocity or astrometric studies and many more are expected from population statistics. The mid-infrared detection of a temperate Super-Jupiter around one of the most close-by solar-type stars Epsilon Indi A using JWST \citep{Matthews2024} is a milestone. Excitingly, its thermal emission at 10 $\mu$m is visible in archival VLT data. Knowing the orders-of-magnitude increase in sensitivity that will be achieved with the ELTs, such a planet can be detected within minutes of observing time. It may already be detectable at $\sim$4 $\mu$m using HRS with current telescopes to characterise the molecular content of its atmosphere, illustrating the enormous potential of the ELTs. 

\subsection{Towards Earth-like planets}

The ultimate goal for the ELTs, and arguably the exoplanet field as a whole, is to characterise Earth-like planets and establish whether they could be habitable and show signs of life. This involves a cascade of questions about temperate rocky planets: Do these planets have an atmosphere? What are their chemical constituents? What are their climates like? Is there any water? Are there biomarker gases present? Are these biomarkers indeed due to biological activity? This final question will be extremely challenging to address. Temperate rocky planets in red dwarf systems are the most accessible for characterisation, and JWST is paving the way by addressing the first and possibly the second question above for a subset of the TRAPPIST-1 planets \citep[e.g.][]{2023Natur.620..746Z}. If their atmospheric conditions are favourable, HRS transmission spectroscopy with the ELT will be sensitive to molecular absorption including molecular oxygen, although the latter will be challenging  \citep{snellen_finding_2013,2014ApJ...781...54R,2023PSJ.....4...83C}.

However, it is high-contrast spectroscopy that will be the real game-changer for the characterisation of other Earths. 
METIS HRS-IFU mode will be able to target Proxima Cen\,b and probe the atmosphere for H$_2$O, CO$_2$, and CH$_4$ in L and M band. If the atmosphere is water-rich, deuterium and the D/H ratio may be determined via HDO measurements at 3.5 $\mu$m \citep{molliere_detecting_2019} and shed light on its history of atmospheric loss. ANDES with its small IFU mode, will determine its albedo over the optical and near-infrared, constrain cloud cover, and probe gaseous species, including molecular oxygen \citep{lovis_atmospheric_2017,palle_ground-breaking_2023,2024MNRAS.528.3509V}. 

The results of the first and second generation of ELT instrumentation will greatly depend on the specific properties of rocky planets around M-dwarfs, such as their geometric albedo (reflected light is expected to dominate over thermal emission at $< 5\mu$m). Results on Proxima\,b will also largely be governed by the contrast limits that can be reached with adaptive optics, which is more challenging at shorter wavelengths. Dedicated instrumentation will be needed to facilitate extreme adaptive optics for the ELTs at the level of the state-of-the-art high-contrast imaging facilities at current 8-10m telescopes. This instrumentation is currently being developed for the European ELT \citep[PCS;][]{2021Msngr.182...38K} and the GMT \citep[GMagAO-X;][]{2022SPIE12185E..4JM}. It is still an open question at what spectral resolution these instruments will operate. \cite{landman_trade-offs_2023} presented detailed simulations for gas giant planets and showed that medium resolution is optimal for this type of planet from a signal-to-noise point of view. However, in the case of reflected light, it is probably necessary to spectrally separate the Doppler shifted planet spectrum from direct starlight to avoid confusion from near-identical telluric and stellar features, pushing requirements to the high-resolution regime. It is envisaged that the high spectral and spatial resolution at extreme contrast levels will eventually allow the characterisation of up to tens of temperate rocky planets around nearby M-dwarfs \citep{2024arXiv240511423H}. 

Current projections suggest that the 10$^{9 - 10}$ contrasts needed to detect and characterise Earth-twins around solar type stars will not be reached with ground-based instrumentations and will require space observatories such as the proposed Habitable Worlds Observatory (HWO\footnote{https://habitableworldsobservatory.org/home}) or a space interferometer at mid-infrared wavelengths (e.g. the LIFE mission\footnote{https://life-space-mission.com}. It has been contemplated whether space-based high-resolution spectroscopy would be an interesting avenue to pursue, knowing the technical challenges. Evidently, the absence of telluric contamination removes an important reason to use HRS. While the dominated noise source for ground-based observations is that from leaked starlight, this is likely to be detector noise in the case of space-based observations. Hence, without new technologies that will significantly reduce detector noise, high spectral resolution (spreading the light over many more detector pixels) is less favourable. Taking these considerations into account it is apparent that R$\sim$100$-$5000 spectroscopy will have the largest discovery space and characterisation power for Earth-twins \citep{wang_observing_2017} using HWO-type space observatories. The power of high-resolution spectroscopy is closely tied to its use with the largest ground-based telescopes.  
 

\section*{DISCLOSURE STATEMENT}
The author is not aware of any affiliations, memberships, funding, or financial holdings that might be perceived as affecting the objectivity of this review. 

\section*{ACKNOWLEDGMENTS}
The author is grateful to all the stimulating discussions with many colleagues. In particular, Michael Line, Matteo Brogi, Jayne Birkby, Paul Molli\`ere, Sid Gandhi, Natalie Grasser, Dario G\'onzalez Picos, Yapeng Zhang, and Sam de Regt are thanked for carefully reading of the manuscript and their comments.
%

\bibliographystyle{ar-style2} 
\bibliography{Library_AllPapers1Apr2024}

\newpage

\vspace{3cm}

\noindent {\Large Supplementary Tables for\\ {\sl Exoplanet Atmospheres at High Spectral Resolution}}\\

\vspace{-3mm}

\noindent Ignas Snellen, Leiden Observatory, Leiden University, \\Postbus 9513, 2300 RA Leiden, The Netherlands\\

\begin{table}[b]
\caption{Data used to produce Fig. 5, showing planet name, apparent V-band magnitude of the host star, equilibrium temperature, atmospheric scale height, Scale Heigh Contrast, amplitude of sodium signal, and the literature references.  All parameters, except sodium signal, are derived from data in the NASA Exoplanet Archive (https://exoplanetarchive.ipac.caltech.edu). Note that the literature measurement of WASP-67 b could not be converted into a transmission signal amplitude.}
\label{AppendixTable1}
\scriptsize
\begin{tabular}{rrrrrrl}\\ \hline
Planet & V$_{\rm{star}}$ & T$_{\rm{eq}}$ & H & SHC &  Na  & Ref.\\
            &   mag             &        K        &    km    &    \%  &    \% & \\      \hline

HAT-P-67 b  & 10.1 &     1930    &   3570  & 0.033    &   $-$&    \citet{bello-arufe_transmission_2023}  \\
HAT-P-70 b  & 9.5&      2550   &   $>$190  &$>$0.0030  &  0.203&  \citet{bello-arufe_mining_2022}     \\
HD 189733 b   &7.7 &     1190   &    200  & 0.011   & 0.39&    \citet{Langeveld_survey_2022}   \\
HD 209458 b   &7.7 &     1480   &    570 &  0.016   & $<$0.41$^\dagger$&   \citet{snellen_ground-based_2008}    \\
KELT-11 b    &8.0 &    1700   &    2630 &  0.014   & 0.5&    \citet{mounzer_hot_2022}   \\
KELT-9 b     & 7.6&   4050   &    730  & 0.0071   & 0.16&    \citet{Langeveld_survey_2022}   \\
MASCARA-2 b   & 7.6&     2260  &    $>$290 & $>$0.0060 &   0.37&    \citet{casasayas-barris_atmospheric_2019}   \\
MASCARA-4 b    & 8.2&    2400  &     260 &  0.0031  &  0.2&    \citet{zhang_transmission_2022}   \\
WASP 21 b     & 11.6&   1260   &    700  & 0.019   &  1.16&      \citet{Langeveld_survey_2022} \\
WASP-121 b   & 10.5&     2450  &     940 &  0.022  &  0.65&    \citet{Langeveld_survey_2022}   \\
WASP-127 b   & 10.1&     1420  &     2150  & 0.046 &   0.34&    \citet{zak_high-resolution_2019}   \\
WASP-166 b   & 9.4&     1270  &     720  & 0.0088  & 0.44&    \citet{seidel_hot_2022}   \\
WASP-17 b   &11.6 &     1700    &   1100 & 0.027    & 1.80&      \citet{zhou_detection_2012} \\
WASP-172 b  & 11.0&      1750 &     1330 &  0.016   & 0.30&     \citet{seidel_detection_2023}  \\
WASP-189 b  & 6.6&      2640  &     500 &  0.0042  &  0.13&    \citet{Langeveld_survey_2022}   \\
WASP-49 b   & 11.4&     1370  &     660  & 0.022   &  2.00&      \citet{wyttenbach_hot_2017} \\
WASP-52 b   & 12.2&     1300  &     660  & 0.039  &  1.30&       \citet{chen_detection_2020-1}\\
WASP-69 b   & 9.9&     990  &     610   &0.026    & 3.28&   \citet{Langeveld_survey_2022}    \\
WASP-7 b   & 9.5&     1490    &   400 &  0.0074   & 0.50&     \citet{rahmati_transmission_2022}  \\
WASP-76 b   & 9.5&     2180   &    1150&   0.020 &   0.47&      \citet{Langeveld_survey_2022} \\
WASP-79 b    & 10.0&    1720  &     680&   0.013   &  1.12&      \citet{Langeveld_survey_2022} \\ \hline

\end{tabular}

$^\dagger$Conflicting non-detection by \citet{casasayas-barris_atmosphere_2021}
\end{table}

\pagebreak

\begin{landscape}

\begin{table}
\caption{Data used to produce Figures 10, 11, and 12, showing the object name, object type (BS = field brown dwarf; YBD = young brown dwarf; SJ = Super Jupiter), C/O ratio, metallicity, $^{12}CO/^{13}CO$ isotope ratio, $v\sin{i}$, and literature references.}
\label{AppendixTable2}
\scriptsize
\begin{tabular}{rrrrrrrl}\\ \hline
Object & Type & Radius & C/O Ratio & C/H & $\frac{^{12}CO}{^{13}CO}$ & $v\sin{i}$ & Ref. \\
            &         & (R$_{\rm{J}})$ &        &        &              &   (km s$^{-1}$) & \\ \hline
DENIS J0255 & BD & 0.78$\pm0.01$ & 0.68$\pm$0.005 & 0.03$\pm$0.05 &184$^{+61}_{-40}$ & 41.0$\pm$0.2 & \citet{deregt2024}  \\
TWA 28 &  YBD & 2.8$\pm$0.2  & 0.61$\pm$0.02 & 0.24$\pm$0.1 & 81$^{+28}_{-19}$ & 11.6$\pm$0.1 & \citet{2024arXiv240707678G}\\
2M J0856& YBD & 1.86$\pm$0.04  & 0.58$\pm$0.01 & 0.26$\pm$0.1 & 79$^{+20}_{-14}$ & 4.5$\pm$0.2& \citet{2024arXiv240707678G}\\
HIP 99770 b & SJ & 1.1$\pm$0.1  & 0.55$\pm$0.05 & 0.26$\pm$0.24 &  & $<$7.8 & \citet{2024AJ....168..131Z}\\
Beta Pic b & SJ & 1.4$\pm$0.1  & 0.41$\pm$0.05 & $-$0.39$\pm$0.16 & & 19.9$\pm$1.0 & \citet{landman_betapic_2024}\\
2M0355 & YBD & 1.15$\pm$0.1  &0.56$\pm$0.02 & 0.21$\pm$0.1 & 108$^{+10}_{-10}$ & $<$4 & \citet{zhang_2M0355}\\
GQ Lup b & SJ & 3.7$\pm$0.7  &  0.50$\pm$0.01 & 0.17$\pm$0.2  &   53$^{+7}_{-6}$ &  5.6$\pm$0.1 & Gonzalez Picos et al., 2024\\
                 &                    &                                &                           &                         &      153$^{+43}_{-31}$   &                      & \citet{2024ApJ...970...71X}\\
YSES 1 b & SJ & 2.8$\pm$0.5  & 0.57$\pm$0.01 & 0.20 $\pm$0.05 & 88$^{+13}_{-13}$ & 5.3$\pm$0.2 & Zhang et al. 2024\\
YSES 1 c & SJ & 1.2$\pm$0.5  & 0.36$\pm$0.15 & $-$0.26$\pm$0.4 & & 11.3$\pm$2.1 & Zhang et al. 2024\\
HD 984 b & YBD  & 1.6$\pm$0.5  & 0.50$\pm$0.01 & $-$0.62$\pm$0.02 & 98$^{+25}_{-20}$ & 12.7$\pm$0.03 &  \citet{Costes2024}\\
HR 8799 d & SJ   &    1.2 $\pm$ 0.1                 &        &     &    & 10.1$\pm$2.8  & \citet{wang_detection_2021}\\
HR 8799 e & SJ    &     1.2 $\pm$0.1                 &      &      &    & 15.0$\pm$2.5  & \citet{wang_detection_2021}\\
HD 4747 B & BD  & 0.82$\pm$0.2  & 0.66$\pm$0.04 &$-$0.10$\pm$0.17 & & 13.4$\pm$1.3 & \citet{2022ApJ...937...54X}\\
HIP 55507 B & BD  & 1.32$\pm$0.2  & 0.67$\pm$0.04 & 0.24$\pm$0.13 & 98$^{+28}_{-22}$ & 5.5$\pm$0.25& \citet{2024ApJ...962...10X}\\
Kappa And b  & SJ        & 1.35$\pm$0.25  & 0.58$\pm$0.05 & $-$0.3$\pm$0.3 && 39.4$\pm$0.9 &\citet{2024ApJ...970...71X}\\
GSC 6214 b  & SJ  & 1.55$\pm$0.25  & 0.70$\pm$0.07 & $-$0.15$\pm$0.4 && 11.6$\pm$2.0 &\citet{2024ApJ...970...71X}\\
ROXs 42B   & SJ          &  2.10$\pm$0.25  & 0.48$\pm$0.08 & 0.0$\pm$0.6 & & 4.4$\pm$2.0 &\citet{2024ApJ...970...71X}\\
DH Tau        & SJ          &2.6$\pm$0.6       & 0.54$\pm$0.06 & $-$0.30$\pm$0.30 & 53$^{+50}_{-24}$ & 5.7$\pm$1.0 &\citet{2024ApJ...970...71X}\\
ROXs 12     & SJ           & 2.2$\pm$0.35   & 0.54$\pm$0.05 &  $-$0.30$\pm$0.30 & & 3.6$\pm$1.5 &\citet{2024ApJ...970...71X}\\
HIP 79098     & SJ       &  2.6$\pm$0.6   & 0.54$\pm$0.03 & $-$0.15$\pm$0.3 & 52$^{+22}_{-17}$ & 4.0$\pm$1.0 &\citet{2024ApJ...970...71X}\\
2M J01225    & SJ      &  1.2$\pm$0.2    & 0.37$\pm$0.08 & $-$0.3$\pm$0.3 & &19.6$\pm$3.0 & \citet{2024ApJ...970...71X}\\
VHS 1256 b & SJ                & 1.27$\pm$0.2 &                                    &                           & & 13.5$\pm$4 & \citet{bryan_constraints_2018}\\
HD 106906 b & SJ                     & 1.49$\pm$0.4 &                                    &                           & & 9.5$\pm$0.2 & \citet{bryan_as_2020}\\
HD 33632A B  &BD                  & 0.78$\pm$0.2 &                    0.58$\pm$0.14         &        0.0$\pm$0.2           & & 53$\pm$3 & \citet{2024ApJ...971....9H} \\
\end{tabular} 
\end{table}
\end{landscape}

\end{document}